\documentclass[opre]{informs3nothing}  

\DoubleSpacedXI 

\usepackage{latexsym, bbm, enumerate, amssymb, amsmath, libertine,tikz}
\usepackage{gensymb}
\usepackage{booktabs}
\usepackage[utf8]{inputenc}
\usepackage[T1]{fontenc}

\usepackage{natbib}
\bibpunct[, ]{(}{)}{,}{a}{}{,}%
\def\bibfont{\small}%
\graphicspath{{./figures/}}

\TheoremsNumberedThrough     
\ECRepeatTheorems

\EquationsNumberedThrough    

\newcommand{\norm}[1]{\left\lVert#1\right\rVert}

\DeclareMathOperator{\Var}{Var}
\DeclareMathOperator{\Cov}{Cov}

\newcommand{\paranth}[1]{\left(#1\right)}
\newcommand{\bracket}[1]{\left[#1\right]}
\newcommand{\curly}[1]{\left\{#1\right\}}

\TITLE{Optimizing Opinions with Stubborn Agents}
\RUNTITLE{Optimizing Opinions with Stubborn Agents}

\RUNAUTHOR{Hunter and Zaman}
\ARTICLEAUTHORS{%
	\AUTHOR{David Scott Hunter}
	\AFF{%
		Operations Research Center, Massachusetts Institute of Technology, 
		77 Massachusetts Ave.,
		Cambridge, MA~  02139, 
		\EMAIL{dshunter@mit.edu}
	}
	\AUTHOR{Tauhid Zaman}
	\AFF{%
		Department of Operations Management, Yale School of Management, Yale University,  
		165 Whitney Ave Ave., 
		New Haven, CT~ 06511,  
		\EMAIL{tauhid.zaman@yale.edu}
	}
} 

\ABSTRACT{
    We consider the problem of optimizing the placement of stubborn agents  in a social network in order to maximally influence the population.   We assume the network contains stubborn users whose opinions do not change, and non-stubborn users who can be persuaded.  We further assume the  opinions in the network are in an equilibrium that is common to many opinion dynamics models, including the well-known DeGroot model.
    
      We develop a discrete optimization formulation for the problem of maximally shifting the equilibrium opinions in a network by targeting users with stubborn agents.  The opinion objective functions we consider are the opinion mean, the opinion variance, and the number of individuals whose opinion exceeds a fixed threshold.  We show that the mean opinion is a monotone submodular function, allowing us to find a good solution using a greedy algorithm.   We find that on real social networks in Twitter consisting of tens of thousands of individuals, a small number of stubborn agents can non-trivially influence the equilibrium opinions.  Furthermore, we show that our greedy algorithm outperforms several common benchmarks.
    
    We then propose an opinion dynamics model where users communicate noisy versions of their opinions, communications are random, users grow more stubborn with time, and there is heterogeneity is how users' stubbornness increases.  We prove that under fairly general conditions on the stubbornness rates of the individuals, the opinions in this model converge to the same equilibrium as the DeGroot model, despite the randomness and user heterogeneity in the model.

}

\begin{document}
\maketitle
\section{Introduction}

Online social networks can be hijacked by malicious actors who run massive influence campaigns on a population and potentially disrupt societies.  Online extremists have utilized social networks to recruit and spread propaganda \citep{markszamanisis}.
There have been multiple reports alleging that foreign actors attempted to penetrate U.S. social networks in order to manipulate elections \citep{ref:russianbots,shane2017fake,guilbeault2016twitter,byrnes2016bot}.  Additionally, there have been other studies suggesting similar operations occurred in European elections \citep{ferrara2017disinformation}. The perpetrators created fake accounts, or ``bots'', 
which shared  politically polarizing content, much of it fake news, in order to amplify it and extend its reach. Furthermore, many of these fake accounts also directly interacted  with humans to promote their agenda \citep{ref:russianbots1}. While no one knows exactly how many people were impacted by these influence campaigns, it has still become a concern for the U.S. government.  Members of Congress have not been satisfied with the response of major  social networks \citep{ref:russianbots_govtresponse} and have asked them to take
actions to prevent future interference in the U.S. democratic process by foreign actors   \citep{ref:russianbots_feinstein}.  

Social network counter-measures are needed to combat these influence campaigns.  This could consist of one using social network agents to influence a network in a way that negates the effect of the malicious influence campaign.  There are multiple components to such counter-measures, but a key one is identifying targets in the network for influence by these agents.  If one has a limited supply of agents, then one needs a method to optimally identify high value targets.

\textbf{Our Contributions.}  In this work we present a method to identify such targets and quantify the impact of \textit{stubborn} agents on the opinions of others in the network. We consider a model for the equilibrium opinion distribution in a network in the presence of stubborn agents whose opinions do not change.  We then present a discrete optimization formulation for the problem of optimally placing stubborn agents in a network to maximally shift the equilibrium opinions.  We consider a slight variant of the traditional influence maximization approaches where instead of converting a non-stubborn agent into a stubborn agent, we simply introduce a stubborn agent into the network and constrain the number of individuals that this agent can communicate with.  We consider three objective functions: the opinion mean, the opinion variance,  and the number of individuals in the network whose opinion is above a given threshold.  We show that the mean opinion is a monotone and submodular function, allowing us to utilize a greedy approach where we have the stubborn agent target individuals one at a time in the network.  

 We show how to apply the greedy algorithm to real social networks.  In particular, we present a neural network based approach to measure opinions and identify stubborn agents.  We find that the opinion equilibrium model is able to predict the true opinions of users in the network with good accuracy, which provides empirical support for its use in opinion optimization.  Using our greedy algorithm, we show that stubborn agents strategically targeting a small number of nodes can non-trivially shift the equilibrium opinions.  Furthermore, we show that our greedy algorithm outperforms several common benchmarks.

 To provide theoretical support for the opinion equilibrium, we  propose a model for opinion dynamics in a social network.  A novel aspect of our model is that the individuals are allowed to  grow stubborn with time and be less affected by new social media posts.  This reflects real behaviors in social networks and is motivated by research in both social psychology and political science. We also allow the stubbornness rate to be heterogeneous across individuals.  We prove that under fairly general conditions on the stubbornness rate, when there are stubborn agents in the network, the opinions converge to the equilibrium condition we use for opinion optimization.

This paper is outlined as follows.  We begin with a literature review in Section \ref{sec:lit_review}. We the present the opinion equilibrium condition in Section \ref{sec:equilibrium}. Our greedy algorithm for stubborn agent placement is presented in Section \ref{sec:optimization}.    Performance results  for our algorithm on real social networks are presented in Section \ref{sec:results}.  We then present our opinion dynamics model in Section \ref{sec:model}. Convergence results for the model are presented in Section \ref{sec:stubborn_agents}. We conclude in Section \ref{sec:conclusion}.  We include all proofs in Section \ref{sec:proofs}, details on the construction of our datasets in \ref{sec:data_analysis}, and implementation details for estimating the polarity of social media posts in Section \ref{sec:neural_network}. 

\section{Literature Review}\label{sec:lit_review}
There has been a rich literature studying opinion dynamics
in social networks.  One of the most popular models here
is the voter model \citep{clifford1973model,holley1975ergodic} where each node updates its opinion to match that of a randomly selected neighbor.
There is a large body of literature studying limiting behavior
in this model \citep{cox1986diffusive, gray1986duality, krapivsky1992kinetics, liggett2012interacting, sood2005voter}.  The model of \cite{degroot1974reaching} is another popular way to describe opinion dynamics.  In this model, a node's opinion is updated to a convex combination of the opinions of itself and its neighbors.   This model has connections with distributed consensus algorithms \citep{tsitsiklis1984problems,tsitsiklis1986distributed,olshevsky2009convergence,jadbabaie2003coordination}.  In contrast to these approaches, there are also Bayesian models of opinion dynamics in social networks \citep{bikhchandani1992theory, banerjee2004word,acemoglu2011bayesian,banerjee1992simple,jackson2010social}.  In these model, a node's opinion is updated using Bayes' Theorem applied to the opinions of its neighbors. 

The notion of stubborn agents 
with immutable opinions was introduced by \cite{mobilia2003does}.  Analysis has been done of the impact of stubborn agents in various opinion models \citep{galam2007role,wu2004social, chinellato2015dynamical, mobilia2007role,yildiz2013binary,acemouglu2013opinion,ghaderi2013opinion}.  It has been shown that strategically placing  stubborn agents in a population can lead to undemocratic outcomes in certain voting schemes \citep{galam2017geometric}.

\cite{ghaderi2013opinion} studied stubborn agents in the model of \cite{degroot1974reaching} and observed that an analogy can be made between the equilibrium opinions and voltages in an electrical circuit.  This electric circuit connection led  \cite{vassio2014message} to propose a function known as harmonic influence centrality, which measured how much a single node could shift the average opinion in the network by switching its own opinion.

The question of optimizing the placement of agents in a social network
to maximize some type of influence was first proposed by \cite{kempe2003maximizing} for a diffusion model. 
Subsequent results have presented a variety of algorithms for this problem \citep{kempe2005influential,leskovec2007cost,chen2009efficient,chen2010scalable}.   \cite{yildiz2013binary} studied optimal stubborn agent placement in the voter model.  Generally speaking, these algorithms make use of the fact that the objective function is submodular, so a good solution can be found using a greedy approach, as shown by \cite{nemhauser1978analysis}.  Our optimization formulation for placing agents in a network also makes use of this property.  

While much analysis has been done on the effect of stubborn agents,
the models used assume that the other individuals in the network
have stationary behavior.  However, numerous psychological studies  have found that people grow stubborn over time (see the review in  \cite{roberts2006patterns} and the references therein). In politics especially, the bulk of empirical evidence supports the hypothesis that susceptibility to changes in ideology and partisanship is very high during early adulthood and significantly lower later in life \citep{alwin1991women,alwin1991aging,sears1975political,sears1981life,sears1983persistence,glenn1980values,jennings1984partisan,jennings2014generations,markus1979political,converse1979plus,sears1999evidence}. Therefore, we believe that opinion dynamics models should include time-varying opinion update processes, where agents become stubborn with time.
Convergence conditions under time-varying dynamics have been studied in \cite{chatterjee1977towards} and later in \cite{hatano2005agreement,wu2006synchronization,tahbaz2008necessary}.  These models do not explicitly consider increasing stubbornness nor the presence of stubborn agents. 
To the best of our knowledge, our work is the first to rigorously analyze convergence in an opinion dynamics model with stubborn agents and increasing stubbornness. 

Additionally, previous opinion dynamics models assume individuals communicate their exact opinion in the network.  However, in reality people may only transmit a noisy version of their latent opinion.
	The previous psychological survey of \cite{mason2007situating}, and the references therein, have argued for modeling latent opinions on a continuous spectrum while allowing for modeling the information communicated between agents on an arbitrary (potentially discrete) spectrum. For instance, we often can only observe an agent's binary decision, but there are frequently many benefits to allowing their underlying latent opinion to be modeled on a continuous spectrum. To the best of our knowledge, \cite{urbig2003attitude} is the only study to consider a framework that separately models communicated and latent opinion, and in this study they do not consider this process with mathematical rigor.


\section{Stubborn Agents and Opinion Equilibrium}\label{sec:equilibrium}
 The core operational equation in this work is an expression for the equilibrium opinions in a social network.  Here we present some notation, the equilibrium equation, and intuition for the equilibrium.  Later in Section \ref{ssec:datasets} we will show empirical evidence that this equilibrium describes opinions observed in real social networks.  To complement this empirical evidence, we will provide theoretical justification for the equilibrium in Section \ref{sec:model}.  In particular, we will show that this equilibrium arises for very general class of models for the dynamics of opinions in a social network.  

\subsection{Notation}\label{sec:notation}
We consider a finite set of agents $\mathcal{V} = \left\{ 1, \ldots, N \right\}$ situated in a social network represented by a directed graph $\mathcal{G} \left(\mathcal{V}, \mathcal{E} \right)$, where $\mathcal{E}$ is the set of edges representing the connectivity among these individuals. An edge $(i, j) \in \mathcal{E}$ is considered to be directed from $i$ to $j$ and this means that agent $i$ can influence agent $j$.  One can view the direction of the edges as indicating the flow of information.  In social networks parlance, we say $j$ \emph{follows}  $i$.  We define the neighbor set of an agent $i \in \mathcal{V}$ as $\mathcal{N}_i = \left\{j \; | \; (j,i) \in \mathcal E \right\}$.  This is the set of individuals who $i$ can be influenced by, i.e. whose posts can  be seen by $i$.  For clarity of exposition, we denote the out-degree neighbor set of an agent $i$ as $\mathcal{N}^o_i = \left\{j \; | \; (i,j) \in \mathcal E \right\}$.  This set is also known as the \emph{followers} of $i$. 

At each time $t\in \mathbb{Z}_{\geq0}$, each agent $i \in \mathcal{V}$ holds an opinion or belief $\theta_i(t) \in [0,1]$.  An opinion near zero indicates opposition to an issue or topic, while an opinion near one indicates support for it.  We define the full vector of opinions at time $t$ by $\theta(t)$ for simplicity. We also allow there to be two types of agents: non-stubborn and stubborn. Non-stubborn agents have an opinion update rule based on communication with their neighbors that we will specify later, while stubborn agents never change their opinions. We will denote the set of stubborn agents by $\mathcal{V}_0 \in \mathcal{V}$ and the set of non-stubborn agents by $\mathcal{V}_1 = \mathcal{V} \; \setminus \; \mathcal{V}_0$. For clarity of exposition, we assume that $\mathcal{V}_0 = \left\{ 1, \ldots, |\mathcal{V}_0 | \right\}$. 

At time $t=0$, each agent $i\in \mathcal{V}$ starts with an initial opinion $\theta_i(0)$. The opinions of the stubborn agents stay constant in time, meaning
\begin{align*}
\theta_i(t) = \theta_i(0), \quad \quad i\in\mathcal{V}_0, \; t\in \mathbb{Z}_{\geq0}.
\end{align*}
To simplify our notation, let $\mathbf \theta_{\mathcal{V}_0}$ denote the vector of the initial opinions of the stubborn agents and $\mathbf \theta_{\mathcal{V}_1}(t)$ denote the vector of the opinions of the non-stubborn agents at time $t$.

\subsection{Opinion Equilibrium}
 The time varying opinions evolve according to an opinion dynamics model.  The most well-known of these models is the DeGroot model where the opinions are deterministic and the update rule for the opinions is linear \citep{degroot1974reaching}.  A key component of the DeGroot model are the influence weights.  We define $p_{ji}$ as the influence weight of agent  $j$ on agent $i$.  In Section \ref{sec:model} we will present an opinion dynamics model where these weights are equal to communication probabilities.  The DeGroot model reaches an equilibrium which is given by a linear system.  This system is characterized by a  $| \mathcal V | \times \mathcal V |$ matrix  $\mathbf A$ given by 
\begin{align}\label{eq:A_matrix}
A_{ik} = \begin{cases}
0 & \text{if } i \in \mathcal{V}_0 \\
-\sum_{j\in \mathcal{N}_i} p_{ji} & \text{if } i \in \mathcal{V}_1, k=i\\
p_{ki} & \text{if } i \in \mathcal{V}_1, k \in \mathcal{N}_i \\
0 & \text{otherwise.}
\end{cases}
\end{align} 
 Throughout, we make the assumption that for all $i\in \mathcal{V}_1$, $\sum_{j\in \mathcal{N}_i} p_{ji} \leq 1$.  
Due to the structure of $\mathbf A$, we can write it in the block-matrix form
\begin{align*}
\mathbf A = \begin{bmatrix}
\mathbf{0} & \mathbf{0} \\
\mathbf{F} &\mathbf{G}
\end{bmatrix}
\end{align*}
where $\mathbf {F}$ is a $|\mathcal{V}_1| \times |\mathcal{V}_0| $ matrix and $\mathbf{G}$ is a $|\mathcal{V}_1| \times |\mathcal{V}_1| $ matrix.  The matrix $\mathbf F$ captures communications from  the stubborn agents to the non-stubborn agents, while $\mathbf G$ captures the communication network among the non-stubborn agents.

We make the assumption that the underlying graph $\mathcal{G} \left(\mathcal{V}, \mathcal{E} \right)$ is connected and for each non-stubborn agent $v\in\mathcal V_1$ there exists a directed path from some stubborn agent to $v$. We note that this assumption is not especially stringent. First off, if the graph has multiple connected components then the results from this section can be applied to each connected component separately. Furthermore, if there are some non-stubborn agents which are not influenced by any stubborn agent, then there is no link in $\mathcal{E}$ connecting the set $\mathcal{R}$ of such non-stubborn agents to $\mathcal{V}\setminus\mathcal{R}$. For these agents there is no unique equilibrium and we do not consider them here.

Let $\theta_i$ be the equilibrium opinion of agent $i\in\mathcal V$ and let $\mathbf \theta_{\mathcal{V}_1}$ denote the vector of equilibrium non-stubborn opinions.  The opinion equilibrium of the DeGroot model is given by the linear system
	\begin{align}\label{eq:equilibrium}
 \mathbf G\mathbf{\theta}_{\mathcal{V}_1} =  \mathbf{F} \mathbf \theta_{\mathcal{V}_0}.
	\end{align}
 From this we see that the non-stubborn opinions are linear combinations of the stubborn opinions.  We can gain more insight into the nature of these linear combinations if examine the equilibrium condition for an individual non-stubborn agent.  Using the expression for $\mathbf A$, we write the equilibrium condition for $i\in\mathcal V_1$ as
\begin{align}
    \theta_i & = \frac{\sum_{j\in\mathcal N_i}p_{ji}\theta_j}{\sum_{j\in\mathcal N_i}p_{ji}}.\label{eq:equilibrium_weighted_avg}
\end{align}
 This expression shows that in equilibrium, the opinion of a non-stubborn agent is a sum of the opinions of those it follows, weighted by their influence weights.

\subsection{Harmonic Influence Centrality}\label{sec:centrality}
The equilibrium condition by equation \eqref{eq:equilibrium}
allows us to  quantify the relative influence of each individual agent in the network.  We define influence as follows.
Imagine we are able to switch an  agent's opinion from zero to one and ask what is the change in the average opinion
in the network as a result of this switch.  This allows us to define \emph{harmonic influence centrality} which was first proposed in \cite{vassio2014message}.  There, harmonic influence centrality measured
how much a stubborn agent increased the average non-stubborn opinion if it flipped its opinion from zero to one while all other stubborn nodes had opinion equal zero.  To make harmonic influence centrality a more operational measure, we consider the actual opinion of stubborn agents in the network rather than setting them all equal to one.  We then define the harmonic influence centrality as  a function $c:\mathcal V\rightarrow\mathbb R$ that maps each
 agent in the network to a real number that equals
the change in average non-stubborn opinion when it is made stubborn and flips its opinion from zero to one.   

 We now present expressions for the harmonic influence centrality of agents
in the network.  To simplify notation we will treat the equilibrium opinions as deterministic. We consider the case of stubborn and non-stubborn agents separately, as they result
in different expressions.  
\begin{theorem}\label{thm:centrality}
	Consider a network with opinion equilibrium given by $-\mathbf G\theta_{\mathcal V_1} = \mathbf F\theta_{\mathcal V_0}$.
	For any stubborn agent $i\in\mathcal V_0$, the harmonic influence centrality is
	\begin{align}
	c(i)& = \frac{-1}{|\mathcal V_1|} \sum_{j\in\mathcal V_1}\paranth{\mathbf G^{-1}\mathbf F}_{ji}\label{eq:centrality_stubborn}
	\end{align}
	and for any non-stubborn agent $i\in\mathcal V_1$, the harmonic influence centrality is
	\begin{align}
	c(i)& = \frac{1}{|\mathcal V_1|-1}\paranth{\frac{\sum_{j\in\mathcal V_1} G^{-1}_{ji}}{ G^{-1}_{ii}}-1}.\label{eq:centrality_nonstubborn}
	\end{align}
\end{theorem}
 We note that our expression for harmonic influence centrality is equivalent to that in \cite{vassio2014message}, with a slight difference in notation.  The authors there defined two matrices $\mathbf Q^{11}$ and $\mathbf Q^{12}$.  In terms of this notation we have $\mathbf F = \mathbf Q^{12}$, and $\mathbf G =\mathbf{Q}^{11} - \mathbf{I}$.  Also, \cite{vassio2014message} defines harmonic influence centrality as the sum of the non-stubborn opinions, whereas we define it as the mean.

The expression for the harmonic influence centrality of stubborn agent $i$ is just the sum
of the $i$th column of the matrix $\mathbf G^{-1}\mathbf F$.
Unlike for stubborn agents, the harmonic influence centrality of non-stubborn agents does not involve the matrix $\mathbf F$
which connects stubborn to non-stubborn agents.  Both expressions require the matrix $\mathbf G$
which connects the non-stubborn agents, to be invertible.  This just means that the network has a unique
opinion equilibrium.  As such, harmonic influence centrality is not applicable in networks where there are no stubborn
agents, or not enough stubborn agents to create a unique equilibrium.  This somewhat limits the applicability
of harmonic influence centrality.  However, it does make its  actual value a relevant operational measure for assessing the influence of individuals in a network.

One useful application of our definition of harmonic influence centrality is in optimizing  functions of equilibrium opinions with stubborn agents. We will see in Section \ref{sec:results} that targeting non-stubborn individuals based on their harmonic influence centrality is a practical and effective approach to impact non-stubborn opinions in very large networks.


\section{Optimization of Stubborn Agent Placement}\label{sec:optimization}
One may be interested in using stubborn agents to shift the opinions in a network in order to maximize a given objective function.  Examples of such functions include the sum of the opinions, the variance of the opinions, or the number of individuals whose opinion exceeds a given threshold.  To optimize these objective functions, we utilize the equilibrium from equation \eqref{eq:equilibrium}.  We now show how to place stubborn agents in a network to shift this equilibrium and optimize these objective functions.

The equilibrium is characterized by a pair of matrices that constrain
the opinions of the non-stubborn individuals.  If stubborn agents are placed in the network, these matrices change. By stubborn agent placement, we mean the agent causes non-stubborn individuals to follow it, allowing the agent to influence their opinion and shift the equilibrium.  By optimizing where we place the stubborn agents, we can shift the opinions in the network as we desire.  

\subsection{Opinion Objective Functions}\label{ssec:objective}
   We now consider the problem of optimizing a function of the non-stubborn equilibrium opinions via stubborn agent placement.  We consider three different objective functions.  First, there is the mean, or equivalently, the sum of the non-stubborn opinions.  This is a fairly standard objective function, and we will see it also has desirable mathematical properties.  Second, there is the number of non-stubborn individuals whose opinion exceeds a given threshold.  This objective function may be relevant if the individuals take an action when their opinion exceeds the threshold (buy a product, vote, protest, etc.).  Third, there is the variance of the opinions.  This objective function can be very important.  For instance, if there is an influence campaign being conducted in a social network to amplify the most extreme opinions, this can increase polarization in the population.  This polarization can lead the loss of faith in institutions and even civil unrest.  To counter this influence campaign, one could to use stubborn agents to decrease this polarization by minimizing the opinion variance.  If instead one wants to amplify polarization in an adversarial population, then one can also maximize the variance.  Minimizing variance is generally a defensive objective, while maximizing variance is more offensive in nature.  Either objective may be of interest to strategic decision makers.

We first consider the sum and threshold objectives.  Consider the scenario where we add one stubborn agent to the network with communication probability $p$. Without loss of generality, we assume that this agent's opinion is one. Suppose that we begin with some equilibrium solution $\theta^0$ that satisfies $-\mathbf{G}\theta^0= \mathbf{F} \theta_{\mathcal{V}_0}$.  Consider adding this new stubborn agent to the network and having non-stubborn individual $i$ follow it. Let $\mathbf{e}_i$ be a vector that has component $i$ equal to one, and all other components equal to zero. When the agent is followed by individual $i$ we  achieve a new equilibrium solution $\theta^1$ given by 
\begin{align*}
-\left(  \mathbf{G} - p\; \mathbf{e}_i\mathbf{e}_i^T \right)\theta^1 = \mathbf{F} \mathbf \theta_{\mathcal{V}_0} + p\mathbf{e}_i.
\end{align*}
The sum of the opinions under this new equilibrium can be written as $ -\mathbf{e}^T \; {\left(\mathbf{G} - p\mathbf{e}_i \mathbf{e}_i^T\right)}^{-1} \left(\mathbf{F} \mathbf \theta_{\mathcal{V}_0} + p \mathbf{e}_i  \right)$, where $\mathbf{e}$ is the vector of all ones.  In general, the stubborn agent can target a set $S\subseteq \mathcal V_1$ of non-stubborn users.  The opinion sum in the resulting equilibrium can be viewed as a function of the target set $S$.  This function $f : \mathcal{V}_1 \mapsto \mathbb{R}_{\geq0}$ is given by 
\begin{align*}
f(S) = -\mathbf{e}^T \; {\left(\mathbf{G} - p \sum_{i\in S}\mathbf{e}_i \mathbf{e}_i^T\right)}^{-1} \left(\mathbf{F} \mathbf \theta_{\mathcal{V}_0} + p   \sum_{i\in S}\mathbf{e}_i  \right).
\end{align*}
In addition to the opinion sum, there are other important functions one can optimize. Consider the set function $g: \mathcal{V}_1 \mapsto \mathbb{R}_{\geq0}$ defined to be 
\begin{align*}
g(S) = \sum\limits_{i\in\mathcal{V}_1} \mathbbm{1}_{i, \tau} \left\{- \; {\left(\mathbf{G} - p \sum_{i\in S}\mathbf{e}_i \mathbf{e}_i^T\right)}^{-1} \left(\mathbf{F} \mathbf \theta_{\mathcal{V}_0} + p   \sum_{i\in S}\mathbf{e}_i  \right) \right\}
\end{align*}
where $\mathbbm{1}_{i, \tau} \left\{\mathbf x \right\}$ is equal to one if the $i$-th component of $\mathbf{x}$ is greater than  some predetermined threshold $\tau$, and zero otherwise. Maximizing this set function is equivalent to maximizing the number of non-stubborn agents with final opinion greater than $\tau$.

 Maximizing variance requires pulling apart the opinions, while minimizing variance requires pushing the opinions together.  One can use two stubborn agents with opposite opinions to accomplish either of these goals.  We can phrase the variance objective in our set function terminology if we redefine the target set.  The target set must indicate which nodes are selected and by which stubborn agents.  Define the stubborn agents as $a_0$ and $a_1$, with opinions zero and one, respectively.  Then define the  set $\mathcal W_1 = \curly{(a,v)|  a\in\curly{a_0,a_1},v\in \mathcal V_i}$.  Each element of $\mathcal W_1$ indicates which stubborn agent targets which non-stubborn node.  A target set for the variance objective will be a subset of $\mathcal W_1$.  To simplify notation, let $|\mathcal V_1|=n$ and let $\theta_i$ be the equilibrium opinion of $i\in\mathcal V_1$ under target set $S\subset\mathcal W_1$.   Then the variance $h: \mathcal{W}_1 \mapsto \mathbb{R}_{\geq0}$ is

\begin{align*}
h(S) = n^{-1}\sum\limits_{i\in\mathcal{V}_1} \theta_i^2 - \paranth{n^{-1}\sum\limits_{i\in\mathcal{V}_1} \theta_i}^2.
\end{align*}

\subsection{Greedy Approach}\label{ssec:algo}
In practice one may limit the number of non-stubborn individuals that are targeted.  This is done so the stubborn agents do not appear to be spam and lose their persuasion power.  A natural constraint would be $|S|=k$ for some $k\leq |\mathcal V_1|$.  Then the problem of determining which $k$ non-stubborn agents to target in order to maximize the sum of the non-stubborn opinions can be written as
\begin{align}\label{eq:optimization_sum}
\max_{S \; : \; S\subseteq \mathcal V_1,|S| = k} f(S).
\end{align}
Similarly, the constrained optimization problem for the number of individuals over a threshold is
\begin{align}\label{eq:optimization_threshold}
\max_{S \; : \;S\subseteq \mathcal V_1, |S| = k} g(S).
\end{align}
 For the variance objective we consider both maximization and minimization.  The corresponding optimization problems are
\begin{align}\label{eq:optimization_var_max}
\max_{S \; : \;S\subseteq \mathcal W_1, |S| = k} h(S),
\end{align}
 and 
\begin{align}\label{eq:optimization_var_min}
\min_{S \; : \; S\subseteq \mathcal W_1,|S| = k} h(S).
\end{align}
These discrete optimization problems become difficult to solve for all $k$ targets simultaneously in real social  networks, which can be quite large.  One solution to this is to solve for one target at a time in a greedy manner.  This means in each iteration choosing the target which gives the largest increase in the objective function.   This  approach greatly reduces the complexity of the problems and allows them to be solved for large networks.  
We next present a set of results concerning the properties of various objective functions.  All proofs can be found in Section \ref{sec:proofs}.

We have a performance guarantee for the greedy approach for the sum of opinions.
\begin{theorem}\label{thm:submodular}
	For an arbitrary instance of $\mathbf{G}$, $\mathbf{F}$, and $\theta_{\mathcal{V}_0}$ the set function $f(\cdot)$ is monotone and submodular.
\end{theorem}
Because the objective is monotone and submodular, a greedy approach to maximizing the sum of opinions  will produce a solution within a factor of $1-e^{-1}$ of the optimum \citep{nemhauser1978analysis}.

 The threshold objective function is not submodular.  Formally, we have the following result.
\begin{theorem}\label{thm:submodular_threshold}
	There exist instances of $\tau$, $\mathbf{G}$, $\mathbf{F}$, and $\theta_{\mathcal{V}_0}$ for which the set function $g(\cdot)$ is not submodular.
\end{theorem}

 The variance objective has a very non-trivial behavior.  We have the following result.

\begin{theorem}\label{thm:variance}
	There exist instances of $\mathbf{G}$, $\mathbf{F}$, and $\theta_{\mathcal{V}_0}$ for which the set function $h(\cdot)$ is not monotone and not submodular.
\end{theorem}

\begin{figure*} 
	\centering
	\includegraphics[scale = .6]{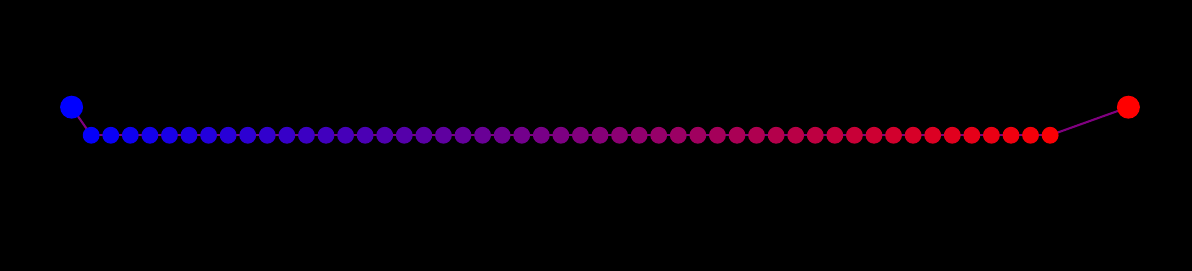}
	\caption{Visualization of an undirected path network with 50 non-stubborn nodes and two stubborn nodes.  The stubborn nodes are located at the ends of the path and have opinions of zero and one.  The nodes are colored by their equilibrium opinions: zero is red and one is blue.  } 
	\label{fig:path} 
\end{figure*}

\begin{figure*} 
	\centering
	\includegraphics[scale = .4]{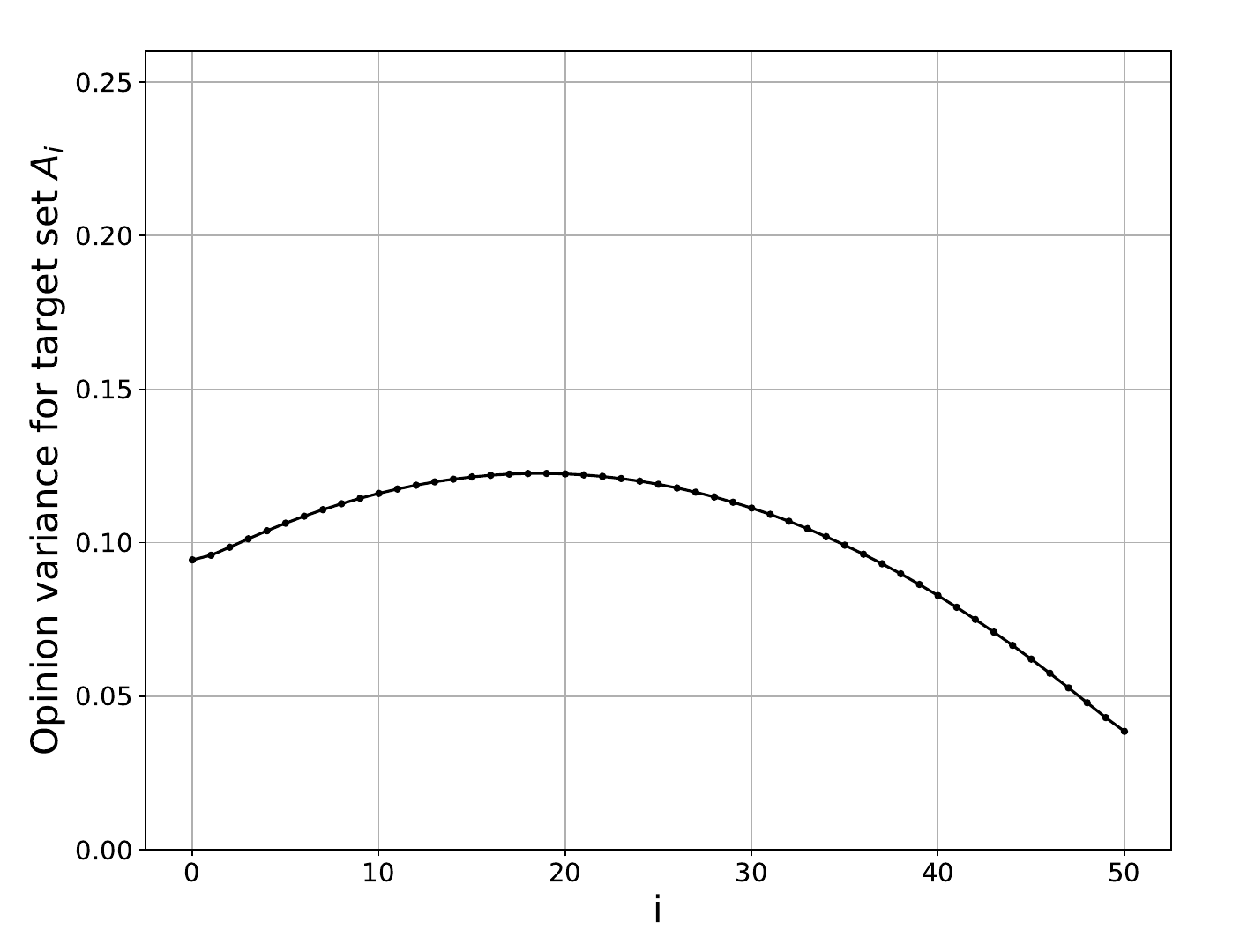}
	\caption{Plot of the non-stubborn opinion variance in the path network from Figure \ref{fig:path} versus the target sets $A_i$ defined in equation \eqref{eq:target_set}.    } 
	\label{fig:variance_nonmonotone} 
\end{figure*}

 For a more visual demonstration of the non-monotone property of the variance objective, consider the network in Figure \ref{fig:path}, which is an undirected path with 50 non-stubborn nodes and
two stubborn nodes.  The stubborn
nodes are located at the ends of the path and have opinions of zero and one.  We assume we have one agent with opinion zero. Let a non-stubborn node a distance $i$ from the stubborn node with opinion zero be denoted node $i-1$.  Define the sets $T_i$ for $0\leq i\leq 49$ as $T_i = \curly{j}_{j=0}^i$.  In Figure \ref{fig:variance_nonmonotone} we plot the opinion variance as a function of $i$, the size of the agent's target set $T_i$.  As can be seen, as the agent targets more nodes the variance increases at first, and then decreases.  What is happening is the first few targets pull the lower opinions closer to zero, which spreads the opinions apart.  Then, as more nodes are targeted by the agent, more opinions are pulled to zero, which decreases the variance.

\subsection{Heuristics for Large Networks}
For very large networks, even our proposed greedy approach for stubborn agent placement remains computationally challenging.  For example, just solving for the equilibrium opinions can take over a second on networks of hundreds of thousands of nodes.  For each iteration of our greedy approach, this equilibrium calculation must be repeated for each potential target node.  For networks with hundreds of thousands of nodes, the time required for each greedy iteration can be on the order of days.  If one wants to find hundreds of targets, the resulting computation can take weeks.  This computational burden can be reduced by not checking every potential target.  However, this could result in severely suboptimal  stubborn agent placement.
To overcome these challenges, we propose some useful computational heuristics.

 First, we describe what we refer to as the \emph{greedy heuristic}. We do not calculate the equilibrium for each potential target.  Instead, we only check a small subset of the targets.  This subset consists of the non-stubborn agents with the highest harmonic influence centrality in the initial network before anyone is targeted. The logic here is that high centrality targets will likely give large gains in the objective functions we consider. In each greedy iteration, we only calculate the equilibrium for the potential targets in this subset.   Calculating harmonic influence centrality  requires solving for the equilibrium twice for each non-stubborn individual in the network. We do not recalculate the centrality in subsequent iterations.

An even simpler heuristic is to target the nodes with the highest harmonic influence centrality.  This approach assumes a sort of independence among the nodes, in the sense that targeting one does not affect the impact of targeting another.

It is useful to compare the computational savings of these heuristics compared to the pure greedy algorithm where every node is checked for each target.  We assume we have a network with $n$ non-stubborn nodes and that we seek $k$ targets for the agents.  For the greedy heuristic, we assume we will only check the top $t$ harmonic influence centrality nodes when searching for targets.  The harmonic influence centrality heuristic requires $2n$ equilibrium calculations.    The greedy heuristic checks $t-i+1$ nodes for the $i$th target.  By summing $i$ from one to $k$, and including the pre-computation of harmonic influence centrality, one finds that the greedy heuristic requires $2n + k(2t-k+1)/2$ equilibrium calculations.  A similar calculation for the pure greedy algorithm shows that it requires  $2n + k(2n-k+1)/2$ equilibrium calculations.  Table \ref{table:compute} summarizes these results.

We can gain insights if we assume that $t,k << n$, which will be true for very large networks. In this case, the greedy heuristic requires approximately $2n+kt$ equilibrium calculations, while the pure greedy algorithm requires $2n+kn$ equilibrium calculations.  The greedy heuristic reduces the number of equilibrium calculations by a factor of $(2n+kn)/(2n+kt)$.  This can be a significant savings for large networks.  For instance, if $n=100,000$, $t = 1,000$, and $k = 1000$, this  reduces the equilibrium computations by a factor of 334.  

Finally, we note that we can parallelize the calculation of the equilibrium.  In each iteration, we can simultaneously calculate the equilibrium of all potential targets.  If enough processors are available the run-time of this step can be reduced to the time for a single equilibrium calculation.  With the resources we had available we were able calculate several hundred equilibria in parallel, increasing our speed by nearly two orders of magnitude.

\begin{table}[!hbt] \centering
	\caption{Number of opinion equilibrium computations needed for different targeting algorithms on a network with $n$ non-stubborn nodes.  We assume $k$ targets are desired and $t$ nodes are checked per target in the greedy heuristic.}
	\label{table:compute}
	\centering
	\begin{tabular}{|c|c|}
		\hline
		Algorithm & Number of opinion equilibrium computations\\\hline
		Greedy &  $2n + k(2n-k+1)/2$\\\hline
	    Greedy heuristic & $2n + k(2t-k+1)/2$\\\hline
		Harmonic influence centrality heuristic &$2n$ \\\hline
	\end{tabular}
\end{table}  


\section{Results}\label{sec:results}
 To understand how much impact a stubborn agent can have on the opinions in a network, we solve the opinion optimization problems described in Section \ref{sec:optimization} with different objective functions on two synthetic networks and two real Twitter social networks.   The synthetic networks are not large in size, but are chosen because they provide insights into how the algorithm selects targets for different objectives.  The Twitter networks allow us to show how to apply the algorithm to real social network data.  This includes training a neural network to measure  opinions, identifying stubborn users for the opinion equilibrium calculation, and scaling up the algorithm for very large networks.  We also use the Twitter networks to show that the greedy heuristic is more effective in optimizing the opinion objective functions than other simpler heuristics.  

\subsection{Performance on Synthetic Networks}
\subsubsection{Path Network}
\begin{figure*} 
	\centering
	\includegraphics[scale = .32]{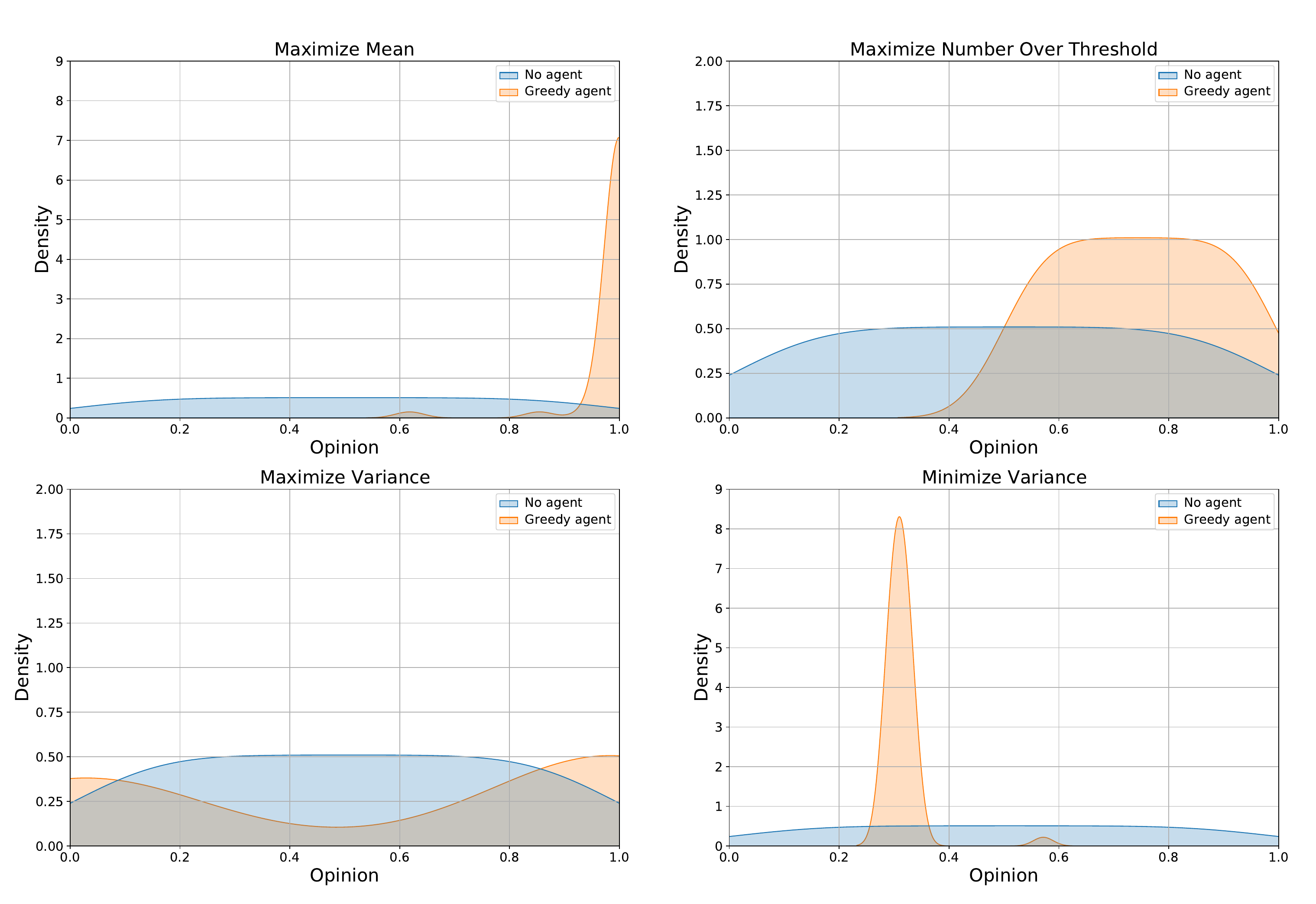}
	\caption{Kernel density estimates for the opinions of non-stubborn nodes in the path network from Figure \ref{fig:path} with and without the greedy agents connected to their targets for each objective function: (top left) maximize mean, (top right) maximize number over a threshold of 0.5, (bottom left) maximize variance, and (bottom right) minimize variance.  } 
	\label{fig:opinion_dist_path} 
\end{figure*}

\begin{figure*} 
	\centering
	\includegraphics[scale = .3]{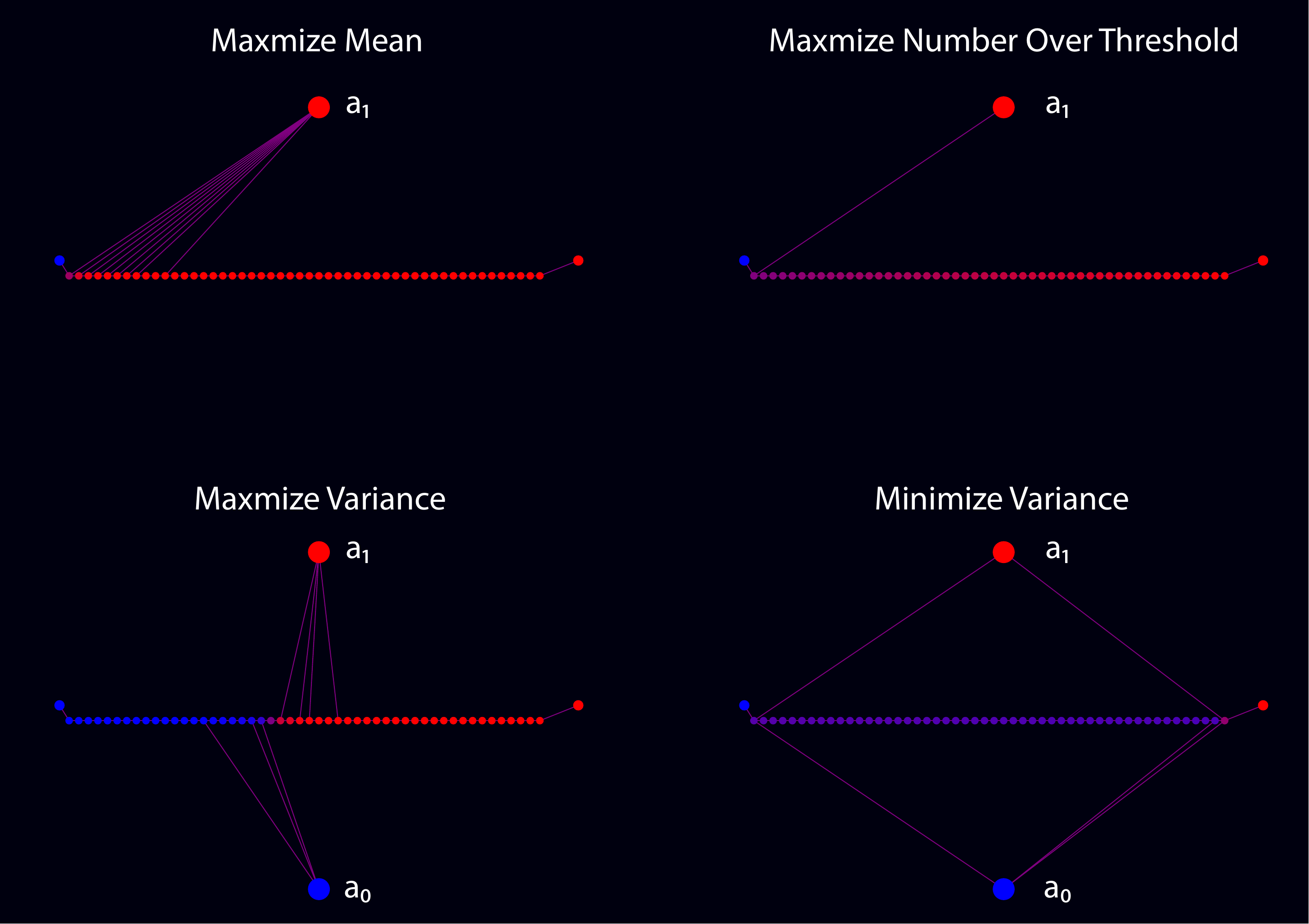}
	\caption{Visualizations of the path network from Figure \ref{fig:path} with the agents $a_0$ and $a_1$ connected to targets based on the greedy algorithm for each objective function: (top left) maximize mean, (top right) maximize number over a threshold of 0.5, (bottom left) maximize variance, and (bottom right) minimize variance.      Each node's color reflects its equilibrium opinion: blue for zero and red for one.  } 
	\label{fig:network_agent} 
\end{figure*}

 We first consider the path network shown in Figure \ref{fig:path}.    We apply the greedy procedure to the four objectives presented in Section \ref{sec:optimization}.  For the threshold objective, we choose a threshold value of 0.5.  We use two stubborn agents $a_0,a_1$ with opinions of zero and one, respectively.  All nodes, including the stubborn agents, have equal communication probability.  We list the initial and final objective values along with the total number of targets in Table \ref{table:obj}.  Kernel density estimates of the non-stubborn opinions with and without our agents are shown in Figure \ref{fig:opinion_dist_path} and the networks with the agents connected to their targets for each objective are shown in Figure \ref{fig:network_agent}.

 We first consider the mean objective.  Only agent $a_1$ chooses targets since we are trying to maximize the mean.  The agent is able to pull the mean very close to one and needs nine targets out of the 50 non-stubborn nodes.  These targets initially have opinions that are close to zero and are all located near the stubborn node with opinion zero.  The resulting opinion distribution, which began uniform, ends up concentrated at one. 

 Next we look at the threshold objective.  Agent $a_1$ needs only a single target to get all 50 non-stubborn nodes above the 0.5 threshold.  The target is the node with the lowest non-stubborn opinion in the network.  Once $a_1$ targets this node, its influence can propagate through the network and get every node above the threshold. 

 To maximize variance, we require two agents.  Agent $a_0$ has three targets and agent $a_1$ has four targets.  From Figure \ref{fig:network_agent} we see that these targets are located in the center of the network.  These nodes initially had opinions near 0.5.  By targeting them, the agents are eliminating these moderate nodes.  The resulting opinion distribution is concentrated at zero and one.  The variance with the agents is 0.23, which is very close to the 0.25 maximum value.

 To minimize variance, we also require two agents.  Agent $a_0$ has two targets and agent $a_1$ has one target.  From Figure \ref{fig:network_agent} we see that these targets are located at the edge of the network and they are targeted by both agents.  These nodes initially had opinions near zero and one.  By targeting them with both agents, these extreme nodes become more moderate.  The resulting opinion distribution is concentrated near 0.3 and  the variance is very close to zero, meaning there is near consensus in the network.

 From these examples we see that the greedy policy selects very different targets depending upon the objective function.  The mean needs many targets because each one pushes the average opinion higher.  However, the threshold objective selects fewer targets because once nodes cross the threshold, there is no value to pushing them any further.  The maximize variance objective targets nodes with moderate opinions and pulls them to the extremes.  The minimize variance objective targets the extreme nodes and pulls them to the middle.  In this instance the network was very balanced in terms of opinion distribution, and so both agents were needed for the two variance objectives.  We will see later that in real networks sometimes only one agent is needed.

\begin{table}[!hbt] \centering
	\caption{Objective function values for the path network from Figure \ref{fig:path} with and without agents, along with the number of targets for each agent.}
	\label{table:obj}
	\centering
	\begin{tabular}{|l|c|c|c|c|}
		\hline
		Objective function & No agents & Greedy agents & Number of targets & Number of targets\\
		                    &  &  & for agent $a_0$ & for agent $a_1$\\\hline
		Maximize mean & 0.500 & 0.988 & 0 & 9\\\hline
		Maximize number over threshold & 25 & 50 & 0 & 1\\\hline
		Maximize variance & 0.082 & 0.232 & 3 & 4\\\hline
		Minimize variance & 0.082 & 0.002 & 2 & 1\\\hline

		\hline
	\end{tabular}
\end{table}  

\subsubsection{Erdos-Renyi Network}
 
We next consider a more complex topology, namely an Erdos-Renyi network with $n$ nodes, where each edge exists independently with probability $p$.  We choose $n = 100$ and $p = 0.1$, which gives each node a mean degree of ten.  The initial opinions are chosen uniformly at random.  We denote two nodes in the network as stubborn with opinions zero and one, respectively.  The stubborn nodes are selected randomly among the nodes in the network.  We again consider two agents $a_0$ and $a_1$ with opinions zero and one, respectively.  We also assume all nodes have equal communication probability.  The number of total targets for the agents is limited to ten.

\begin{figure*} 
	\centering
	\includegraphics[scale = .32]{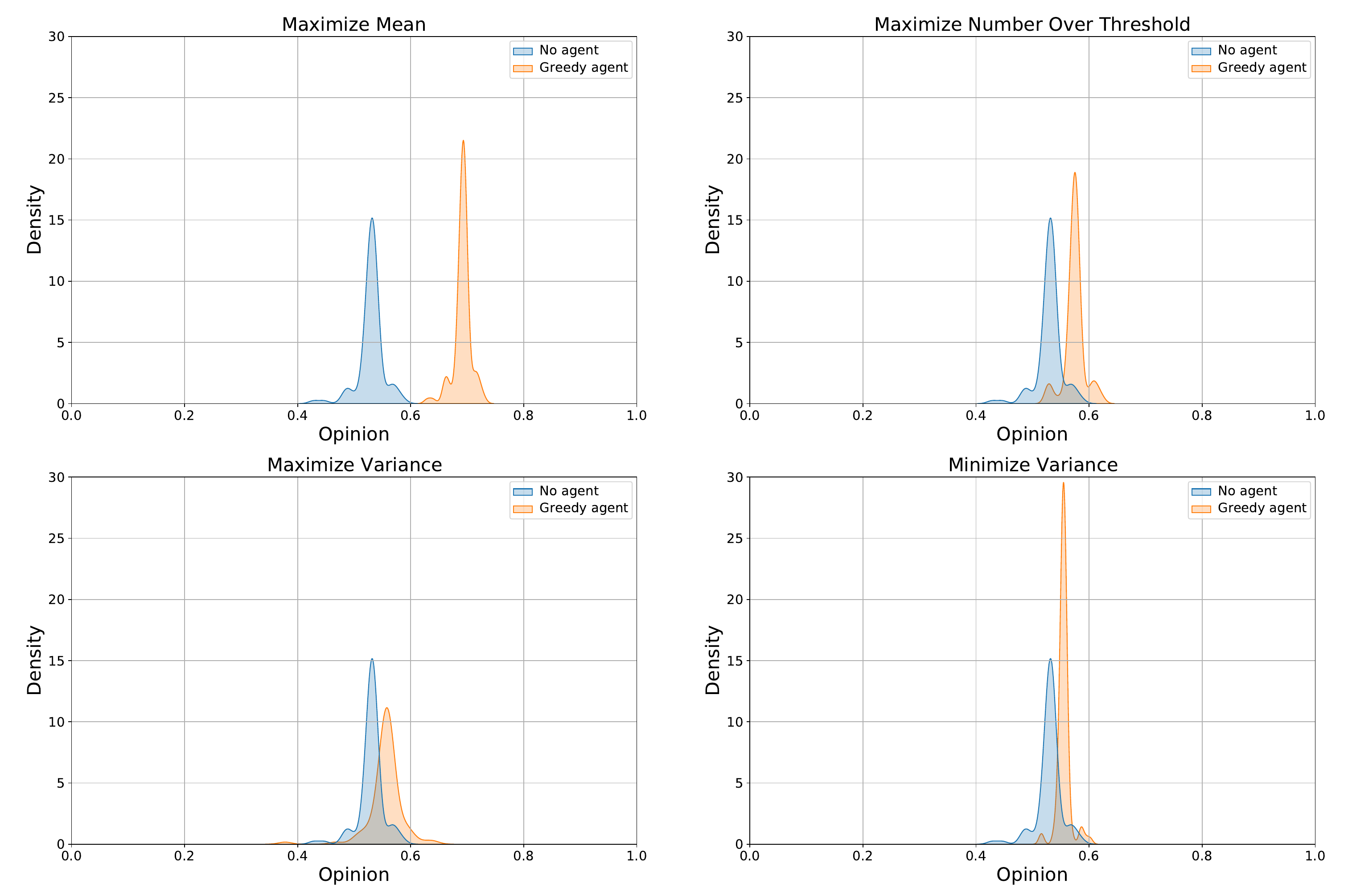}
	\caption{Kernel density estimates for the opinions of non-stubborn nodes in the Erdos-Renyi network with and without the greedy agents connected to their targets for each objective function: (top left) maximize mean, (top right) maximize number over a threshold of 0.5, (bottom left) maximize variance, and (bottom right) minimize variance.  } 
	\label{fig:opinion_dist_ER} 
\end{figure*}

Table \ref{table:obj_ER} shows the values for the objective functions with and without the optimized agents, along with the number of targets for each agent.  We see that each objective requires a different number of targets for the two agents, similar to the path network.  Figure \ref{fig:opinion_dist_ER} shows the distribution of the non-stubborn opinions with and without the agents.  As with Figure \ref{fig:opinion_dist_path}, we see that the distribution with the agents varies with the objective.  The mean objective has the agents pull the opinions near 0.7, while the threshold objective only pulls them near 0.6.  For the variance objective, we see that to maximize it, the agents pull the higher opinions up slightly, creating a wider distribution.  To minimize the variance, the agents pull the lower opinions up to create a narrow distribution.

\begin{table}[!hbt] \centering
	\caption{Objective function values for the Erdos-Renyi network with $n=100$ and $p = 0.1$ with and without agents, along with the number of targets for each agent.}
	\label{table:obj_ER}
	\centering
	\begin{tabular}{|l|c|c|c|c|}
		\hline
		Objective function & No agents & Greedy agents & Number of targets & Number of targets\\
		                    &  &  & for agent $a_0$ & for agent $a_1$\\\hline
		Maximize mean & 0.530 & 0.692 & 0 & 9\\\hline
		Maximize number over threshold & 92 & 100 & 0 & 2\\\hline
		Maximize variance & 0.0005 & 0.0012 & 1 & 2\\\hline
		Minimize variance & 0.0005 & 0.0001 & 3 & 6\\\hline

		\hline
	\end{tabular}
\end{table}  

\subsection{Twitter Datasets}\label{ssec:datasets}
We consider two datasets from the social network Twitter about certain geo-political events.  The first dataset consists of Twitter users discussing Brexit, the planned British departure from the European Union.  The second consists of Twitter users discussing the Gilets Jaunes protests in France.  We chose these events because there may be interest in shifting opinions on these events given their significance.  We now provide some background about these events.

\textbf{Brexit.} Brexit is the withdrawal process of the United Kingdom (UK) from European Union (EU). While Brexit began with a vote in 2016, in this work, we focus on the time period from September 2018 to March 2019 when the British government worked on constructing a formal plan for executing the Brexit.

\textbf{Gilets Jaunes.} Gilets Jaunes, or Yellow Vests, is a French populist movement that started in November 2018. Although it was initially a response to the sudden rise in fuel prices, it quickly became a generalized social unrest protest against the government of president Emmanuel Macron.  The protests have been going on every Saturday since November 2018, each week being called as a new ``Acte'' by the protesters.  In this work, we focus on social network data about the Gilets Jaunes protests from February 2019 to May 2019.

For each event, we identified a set of relevant keywords.  We then collected every post or \emph{tweet}, on the social network site Twitter containing these keywords during the relevant collection periods.  We also collected the follower edges between all users who posted these tweets for each event.  This provided us with the follower network of Twitter users discussing each event.  In addition we were able to measure the posting rate for each user by counting the number of tweets they posted during the data collection period.  We provide basic information about the datasets in Table \ref{table:NNdata_stat}.  Further details on the dataset construction is provided in Section \ref{sec:data_analysis}.
\begin{table}[!hbt] \centering
	\caption{Basic information about the Twitter datasets (M denotes million).}
	\label{table:NNdata_stat}
	\centering
	\begin{tabular}{|l|l|c|c|c|}
		\hline
		Event     &    Data collection   & Number of  & Number of & Number of\\
		          &    period            &  tweets & follower edges & users\\
		\hline
		Brexit 	 & September 2018 to February 2019 &  27M  & 18.5M &104,755 \\ \hline
		Gilets Jaunes  & 	February 2019 to May 2019 & 3.2M & 2.3M & 40,456 \\
		\hline
	\end{tabular}
\end{table}  

\subsection{Opinions and Stubborn Users}\label{sec:data_stubborn}
To apply our equilibrium model, we require the follower network, posting rate, and stubborn user opinions.  We already have the first two items form the raw data.  Next we must identify the stubborn users. We assessed stubbornness using the content of the tweets of the users.  This was done by building a neural network to measure the tweet opinions.  To do this we developed a novel approach to label tweets with a ground truth opinion in order to construct a training dataset for the neural network.
We first identified several extremely polarized hashtags for each event. These are words
or phrases that exhibit strong support for or opposition to the event.  
The complete lists of hashtags for Brexit and Gilets Jaunes are found Section \ref{sec:data_analysis}.  Human domain experts who manually inspected several hundred random user profiles identified these hashtags.  If a hashtag appeared often, was related to the topics, and was politically charged, it was included in the polarized hashtag list.  These hashtags were labeled as either pro-event or anti-event by the domain experts.   We then identified all users in our dataset that had the hashtags in their profile description. If the user is using any of the hashtags from the pro-event list, and none from the anti-event list, then this user is labeled as pro-event.  The same process is done for the anti-event list.  After this user labeling is done, all tweets in the dataset belonging to any anti-event users are given an opinion of zero, and all tweets of pro-event users are given an opinion of one.  The logic here is that if a user puts the extremely polarized hashtags in their profile description, they are broadcasting a very strong signal about their opinion.  It is then highly likely that any tweet they post about the event will have a very extreme opinion.  Using our approach, we are able to efficiently label hundreds of thousands of tweets for the two events.   Details of the training data are provided in Table \ref{table:training_set}.
\begin{table}[!hbt] \centering
	\caption{Details on the neural network training data  for Brexit and Gilets Jaunes. The number of tweets in the training data and number of users who posted these tweets  for and against each event are shown.}
	\label{table:training_set}
	\centering
	\begin{tabular}{|l|c|c|c|c|}
		\hline
		Event &Pro-event tweets &	Anti-event tweets & Pro-event users  	& 	Anti-event users  \\ \hline
		Brexit 		& 	400,000	& 	400,000	& 1,935	&	6,863	\\ \hline
		Gilets Jaunes & 	130,000	& 	130,000	&  383 & 2,354 \\ \hline 
	\end{tabular}
\end{table}

Once we had labeled the tweets, we could train the neural network.  We use a standard architecture that was developed in \cite{kim2014convolutional}. For each event we train on 80\% of the labeled data and tested on the remaining fraction. We used the deep learning library \emph{Keras} \citep{chollet2015}, and trained our model with a cross-entropy loss over five epochs on a single CPU. With this configuration, the training time is under a few hours.  The resulting performance is quite good.  The neural network achieves an accuracy of 86\% on the testing data for Brexit and 83\% on the testing data for Gilets Jaunes. For a more qualitative demonstration of the accuracy of the neural network, we show in  Tables \ref{table:exTweets_BREXIT} and \ref{table:exTweets_YELLOW_VESTS} the opinions it measures for tweets in the Brexit and Gilets Jaunes datasets, respectively.  As can be seen, the opinion estimates of the neural network align with the text of the tweets.  Details of the neural network architecture and training process  are provided in Section \ref{sec:neural_network}.

\begin{table}[h!]
	\begin{center}
		\caption{Tweets and their opinion scores given by the neural network for the Brexit dataset.  An opinion of zero is anti-Brexit and an opinion of one is pro-Brexit.}
		\label{table:exTweets_BREXIT}
		\begin{tabular}{|p{4in}|c|} 
			\hline
		      \multicolumn{1}{|c|}{Tweet} & Polarity  \\	\hline
		    \#stopbrexit \#PeoplesVoter\#brexit \#Eunurses \#nurseshortage & 0.03 \\ \hline
			Britain will receive an economic boost on the back &\\ 
			of a Brexit deal with the European Union, Philip Hammond &\\
			has again claimed & 0.63  \\ \hline
			@Nigel\_Farage Wait for the remoaners to make stupid comments&\\
			 of Russian interference on Brexit &  0.76\\ \hline 
		\end{tabular}
	\end{center}
\end{table}

\begin{table}[h!]
	\begin{center}
		\caption{Tweets and their opinion scores given by the neural network for the Gilets Jaunes dataset.
		An opinion of zero is anti-Gilets Jaunes and an opinion of one is pro-Gilets Jaunes.}
		\label{table:exTweets_YELLOW_VESTS}
		\begin{tabular}{|p{4in}|c|} 
			\hline
			\multicolumn{1}{|c|}{Tweet} & Polarity  \\	\hline
			Il n'y a aucune raison que leurs revendications passent& \\ 
			avant d'autres, quelques dizaines de milliers repr\'esentant & \\
			une minorit\'e ne vont pas d\'ecider pour la majorit\'e. & 0.0\\ \hline
			\#Giletsjaunes \#Nancy Les manifestants ont r\`eussi \`a & \\
			entrer dans le périmètre interdit dans le centre ville. &  0.5 \\ \hline
			Aucun essoufflement pour l'\#ActeXV des \#GiletsJaune! & 0.85 \\\hline 
		\end{tabular}
	\end{center}
\end{table}
Our final step was to identify which users were stubborn based on their opinions.
We used the trained neural network to estimate the opinion of all tweets in 
our datasets.  Then we averaged the opinions of each user's tweets
to obtain their opinions.  
  We determined which users were stubborn by setting lower and upper opinion intervals.  Any user whose opinion falls within either of these intervals is declared stubborn. 
We made the assumption that people with more extreme opinions (close to zero or one) are stubborn.  Previous work in opinion dynamics supports this definition of stubborn. For example,  \cite{martins2013building} define an \emph{inflexible agent} as someone who has a very strong, extreme opinion. Further evidence is provided by \cite{moussaid2013social} who found that the majority of people systematically keep their opinion when their own confidence exceeds that of their partner.  People with extreme opinions are generally confident in their beliefs.  This suggests that people with more extreme opinions are likely to be stubborn.

For our datasets, we chose $[0.0,0.1]$ and $[0.9,1.0]$ as the stubborn intervals.  We performed robustness checks and found that our opinion optimization results were not sensitive to the precise values of these intervals, as long as the values were reasonable and left sufficient non-stubborn users in the network.  Using these stubborn intervals, we have 81,043 non-stubborn users and 23,705 stubborn users for Brexit and 38,483 non-stubborn users and 1,973 stubborn users for Gilets Jaunes. For Brexit there are 6,147 users in $[0.0,0.1]$ and 1,555 users  in $[0.9,1.0]$.  For Gilets Jaunes  there are 1,973 users in $[0.0,0.1]$ and only 134 users in $[0.9,1.0]$.

\begin{table}[h!]
	\begin{center}
		\caption{Number of stubborn and non-stubborn users in the Brexit and Gilets Jaunes datasets.}
		\label{table:stubnotstub_stats}
		\begin{tabular}{|l|c|c|} 
			\hline
			Dataset 						    &  Brexit &   Gilets Jaunes\\ \hline
			Number of non-stubborn users  		  	& 81,043 & 38,483  \\ \hline
			Number of stubborn users  					 & 23,705 & 1,973\\ \hline
			Number of stubborn users in $[0.9,1.0]$   & 5,893 & 134\\ \hline
		Number of stubborn users in $[0.0,0.1]$    & 14,950 & 1,839\\ \hline

		\end{tabular}
	\end{center}
\end{table} 

 Once we establish which users are stubborn, we can use the equilibrium condition (equation \ref{eq:equilibrium}) to predict the opinion of non-stubborn users.  We use the posting rate of the individuals for the influence weight/communication probabilities.  By comparing these equilibrium opinions with the opinions based on the neural network, we can assess how well the opinion equilibrium reflects reality.  

 We solve for the equilibrium opinions for both networks and calculate the correlation coefficient between the equilibrium and neural network opinions.  The values are shown in Table \ref{table:equilibrium_correlation}.  We find that the correlation is 0.78 for both networks.  This value is impressive when one considers that the predictions are made using the opinions of stubborn users, who are a small percentage of the network (22.6\% for Brexit and 4.9\% for Gilets Jaunes).   These results suggest that the opinion equilibrium is effective at characterizing the opinion distribution in real social networks.  

\begin{table}[h!]
	\begin{center}
		\caption{Correlation coefficient of neural network based and equilibrium based non-stubborn opinions.}
		\label{table:equilibrium_correlation}
		\begin{tabular}{|c|c|c|c|} 
			\hline
			Dataset &   Correlation coefficient (p-value)\\
		&  of neural network and equilibrium  opinions \\ \hline
		 Brexit &  0.78 ($<10^{-6}$)\\ \hline
		 Gilets Jaunes & 0.78 ($<10^{-6}$)\\ \hline
		\end{tabular}
	\end{center}
\end{table}

\subsection{Performance on Twitter Datasets}
We applied the greedy heuristic from Section \ref{ssec:algo} to target non-stubborn individuals in the Brexit and Gilets Jaunes networks in order to maximize the mean opinion and the number of individuals with opinion greater than 0.5,  and also to maximize and minimize the opinion variance.    The opinions we are optimizing are calculated from the equilibrium condition in equation \eqref{eq:equilibrium}, using the neural network opinions for the stubborn opinions and posting rates for the influence weights/communication probabilities. For reference, we compared the greedy heuristic to a set of benchmark targeting heuristics which we now describe. 

\begin{itemize}
	\item[\textbf{Out-degree.}] This heuristic targets the nodes in order of decreasing out-degree (follower count).  
	
	\item[\textbf{Posting rate.}]  This heuristic targets the nodes in order of decreasing posting rate.  
		
	\item[\textbf{Harmonic influence centrality.}]  This  heuristic targets the nodes in order of decreasing harmonic influence centrality. 
\end{itemize}  
Each of these benchmark heuristics exploit a different aspect of the opinion equilibrium.  Out-degree focuses on the sum over neighbors.  The logic here is that users with  many followers have more influence.  Posting rate  focuses on the communication probability/rate terms.  The logic here is that users' opinions will tend to align with their active neighbors.  Harmonic influence centrality  combines these two aspects to identify active users with a large reach.  

For the greedy heuristic, we pre-computed the harmonic influence centrality of all non-stubborn users in each network, as mentioned in Section \ref{ssec:algo}.  We then checked 1,000 users with the highest harmonic influence centrality as potential targets in each iteration of the algorithm.

We had the stubborn agent  post at the average rate as the non-stubborn users in the network.  This would prevent the agent from appearing too suspicious and potentially being flagged by spam detection algorithms on the social network.  

\begin{figure*} 
	\centering
	\includegraphics[scale = .6]{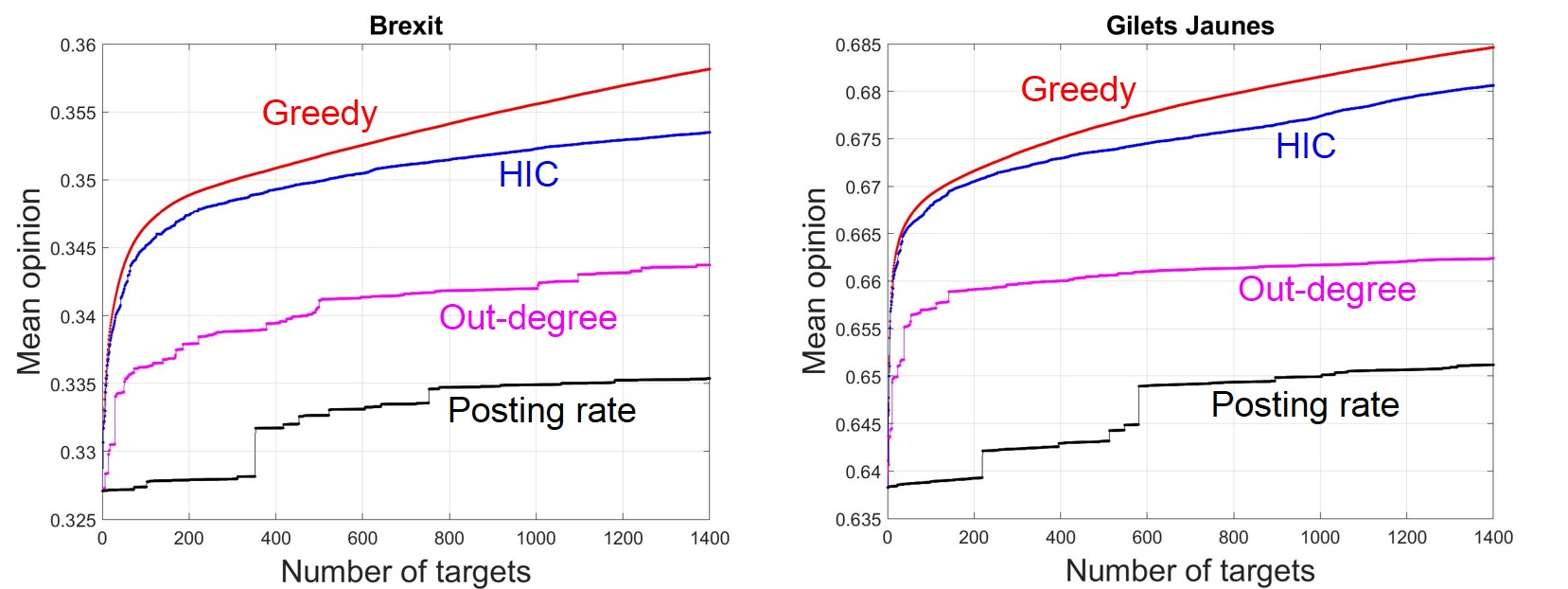}
	\caption{Plot of mean non-stubborn opinion versus number of  targeted non-stubborn individuals in the Brexit network (left) and Gilets Jaunes network (right) for different targeting algorithms (HIC is harmonic influence centrality).  The stubborn agent posts at a rate equal to the mean rate of the network.  } 
	\label{fig:mean_mean} 
\end{figure*}

\begin{figure*} 
	\centering
	\includegraphics[scale = .6]{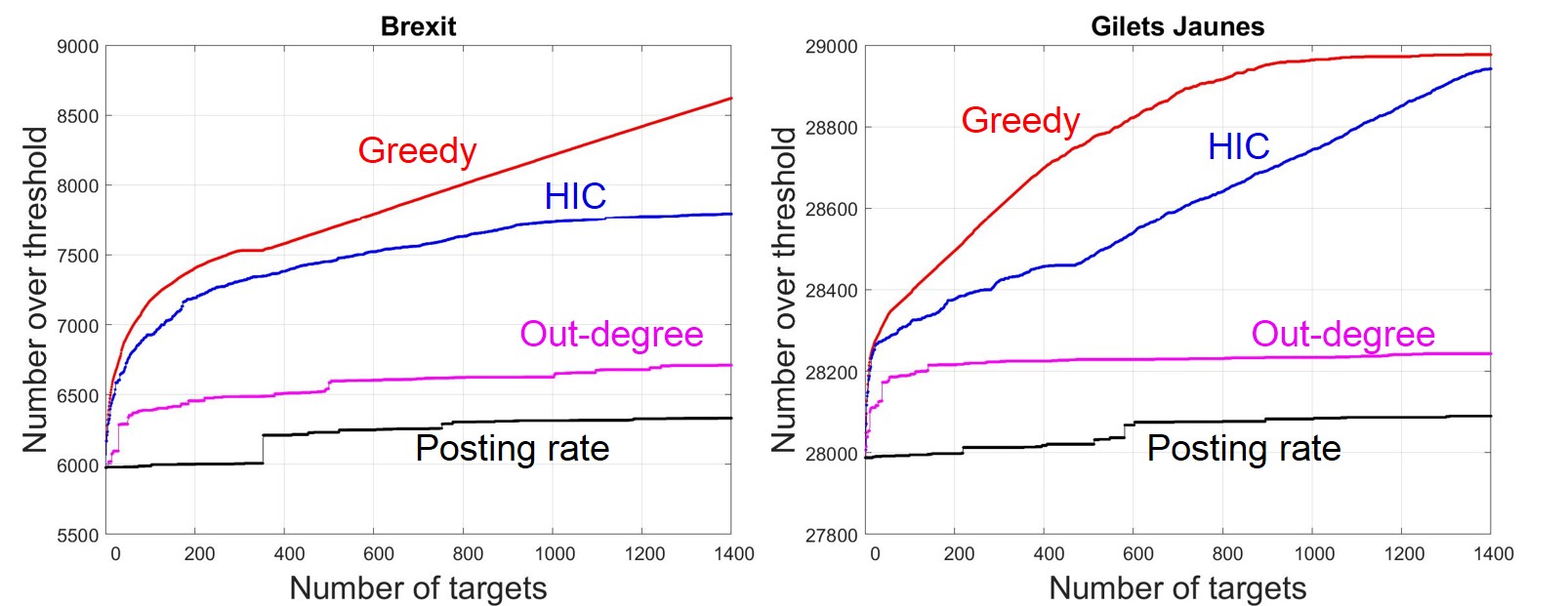}
	\caption{Plot of number of non-stubborn individuals with opinion over the 0.5 threshold versus number of  targeted non-stubborn individuals in the Brexit network (left) and Gilets Jaunes network (right) for different targeting algorithms (HIC is harmonic influence centrality).  The stubborn agent posts at a rate equal to the mean rate of the network.  } 
	\label{fig:threshold_mean} 
\end{figure*}

We start by looking at maximizing the mean and threshold objectives.
Each algorithm's performance for the different networks and objective functions is shown in Figures  \ref{fig:mean_mean} and \ref{fig:threshold_mean}.  We see similar trends for all the scenarios.  The posting rate is the worst algorithm and  harmonic influence centrality does better than out-degree.  Our greedy algorithm has the best performance, which shows the importance of the network structure in the targeting process.  For the mean objective function we see a rapid increase in the mean opinion when less than 100 users are targeted, after which we see a linear growth in the opinion.  This is interesting because it suggests a few targets can have a large impact on the opinions.

For the threshold objective function, we see different results for the two networks.  In Brexit, the greedy algorithm increases the objective function at  a near linear rate as more users are targeted, while harmonic influence centrality saturates.  However, in Gilets Jaunes, the greedy algorithm initially outperforms harmonic influence centrality, but then saturates as more targets are added.  In contrast, harmonic influence centrality steadily increases at a linear rate, eventually catching up with the greedy algorithm.  For Gilets Jaunes, we see that targeting 100 users with the greedy algorithm moves 403 users over the threshold.  In Brexit, targeting 100 users with  the greedy algorithm puts 1,197 users over the threshold.  We see a greater efficiency of the targeting in Brexit compared to Gilets Jaunes.  This may be due to the initial opinion distribution.  For Brexit, there are initially approximately 6,000 non-stubborn users above the threshold, which is 7.4\% of the non-stubborn users.  Therefore, there are many people available to be pushed over the threshold.  In contrast, the Gilets Jaunes network there are initially about 28,000 non-stubborn users above the threshold, or 73.7\% of the non-stubborn users.  In this case, there are fewer people available to be pushed over the threshold.  We suspect this is the reason the Brexit targeting is more efficient.

 The variance objective is not  monotone and not submodular.  Also, it requires us to test each (agent,target) pair in each greedy iteration.  Therefore, we test maximizing and minimizing the variance on smaller sub-networks of the Brexit and Gilets Jaunes datasets so we can test all nodes in each iteration.  These sub-networks have 2,000 nodes which are selected at random.  We use two agents, $a_0$ and $a_1$ with opinions zero and one, respectively. 

\begin{figure*} 
	\centering
	\includegraphics[scale = .3]{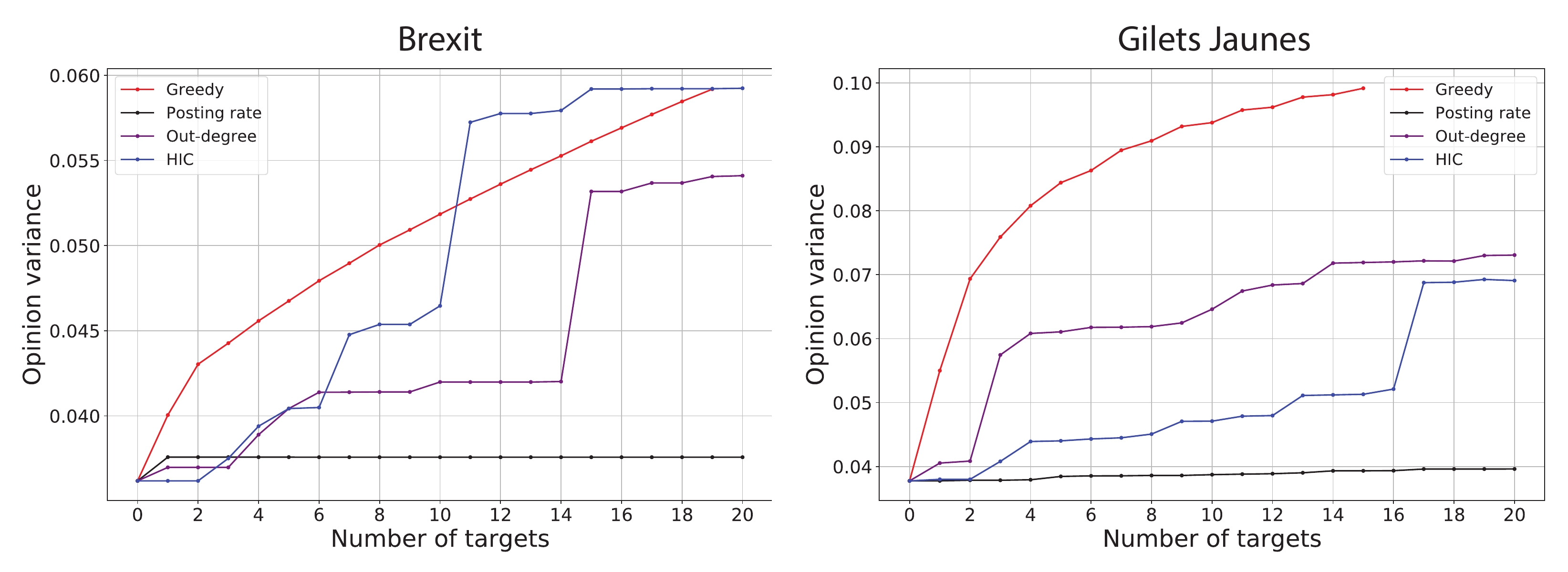}
	\caption{Plot of the variance of non-stubborn opinions versus number of  targeted non-stubborn individuals in in the Brexit and Gilets Jaunes sub-networks for different targeting algorithms (HIC is harmonic influence centrality) trying to maximize the variance.  The stubborn agent posts at a rate equal to the mean rate of the network.   }
	\label{fig:var_max_ntargets} 
\end{figure*}

 We begin by maximizing the variance.  The resulting variance versus number of targets is shown in Figure \ref{fig:var_max_ntargets}.  We also compare to the baseline posting rate, out-degree, and harmonic influence centrality algorithms.  For these algorithms we try each agent with the target and keep the agent that produces the larger variance.  We find that for Gilets Jaunes the greedy procedure is by far the best.  However, for Brexit, harmonic influence centrality outperforms greedy when there are sufficient targets chosen.  The agents can nearly double the variance in both networks.

\begin{figure*} 
	\centering
	\includegraphics[scale = .3]{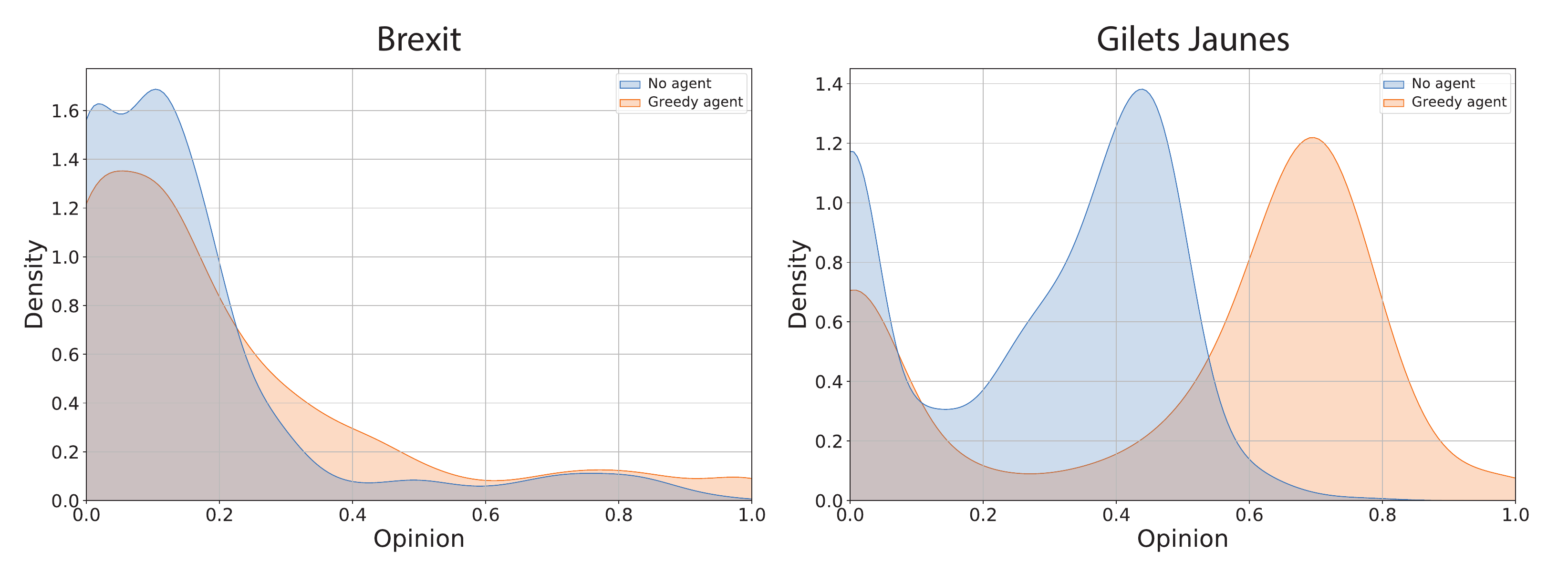}
	\caption{Kernel density estimates for the opinions of non-stubborn nodes in the Brexit and Gilets Jaunes sub-networks with and without the greedy agents connected to their targets. The agents choose targets to maximize the variance.  }
	\label{fig:var_max_opinion_dist} 
\end{figure*}

 We note that in each greedy iteration, the agent chosen was $a_1$ for each network.  To understand why, we look at kernel density estimates for the non-stubborn opinions with and without the agents  in Figure \ref{fig:var_max_opinion_dist}.
For Gilets Jaunes, the opinions without the agent have modes near zero and 0.4.  With the greedy agent connected to the targets, the upper mode shifts near 0.7.  To increase the variance, the greedy algorithm is pulling the upper mode further up.  This is why agent $a_1$ is the selected agent in each iteration.  For Brexit we see a similar phenomenon, but less dramatic in effect.  The agent pulls more opinions into the upper tail of the opinion distribution.  Note that this behavior is unlike the path network in Figure \ref{fig:path}, where we saw targets assigned to both agents.  The initial distribution of the opinions likely effects which agents are chosen.

\begin{figure*} 
	\centering
	\includegraphics[scale = .3]{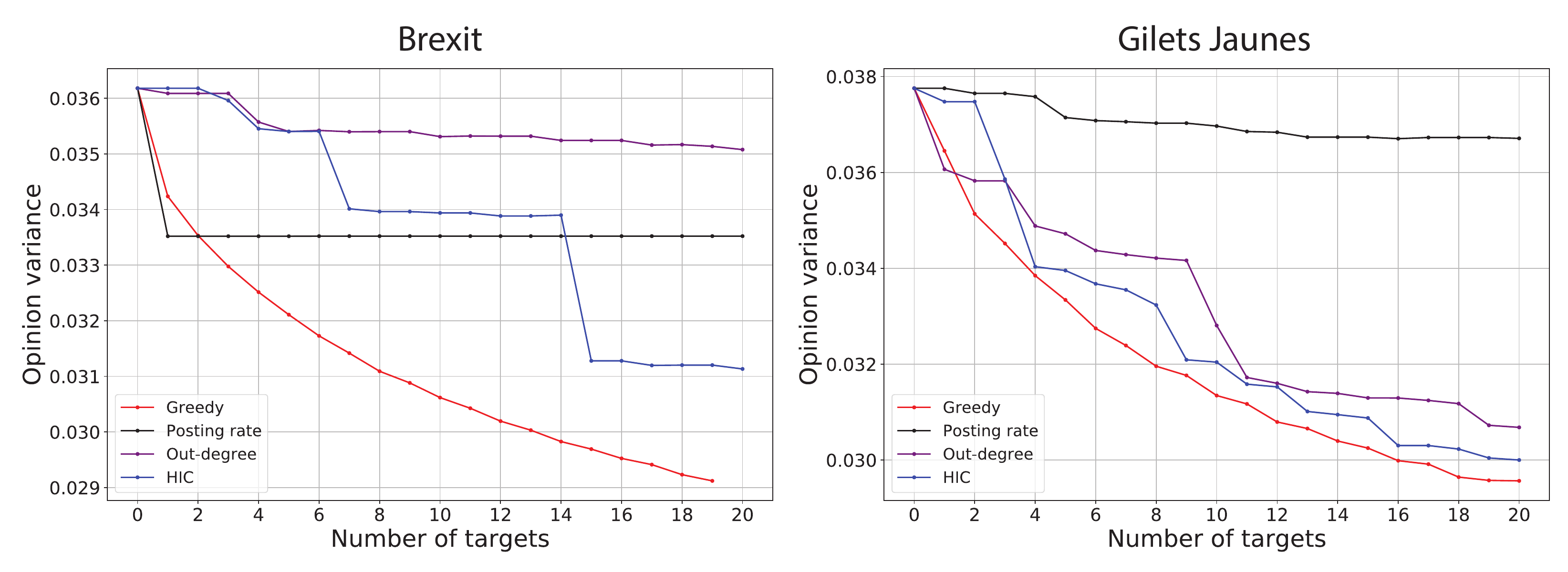}
	\caption{Plot of the variance of non-stubborn opinions versus number of  targeted non-stubborn individuals in in the Brexit and Gilets Jaunes sub-networks for different targeting algorithms (HIC is harmonic influence centrality) trying to minimize the variance.  The stubborn agent posts at a rate equal to the mean rate of the network.   }
	\label{fig:var_min_ntargets} 
\end{figure*}

 The performance for the minimizing variance objective is shown in Figure \ref{fig:var_min_ntargets}.  Unlike with maximizing variance, here we see greedy outperforming all benchmarks for both datasets.  Greedy is much better than the benchmarks on Brexit, while it is only slightly better on Gilets Jaunes. Harmonic influence centrality is the best of the benchmark algorithms on both datasets.

\begin{figure*} 
	\centering
	\includegraphics[scale = .3]{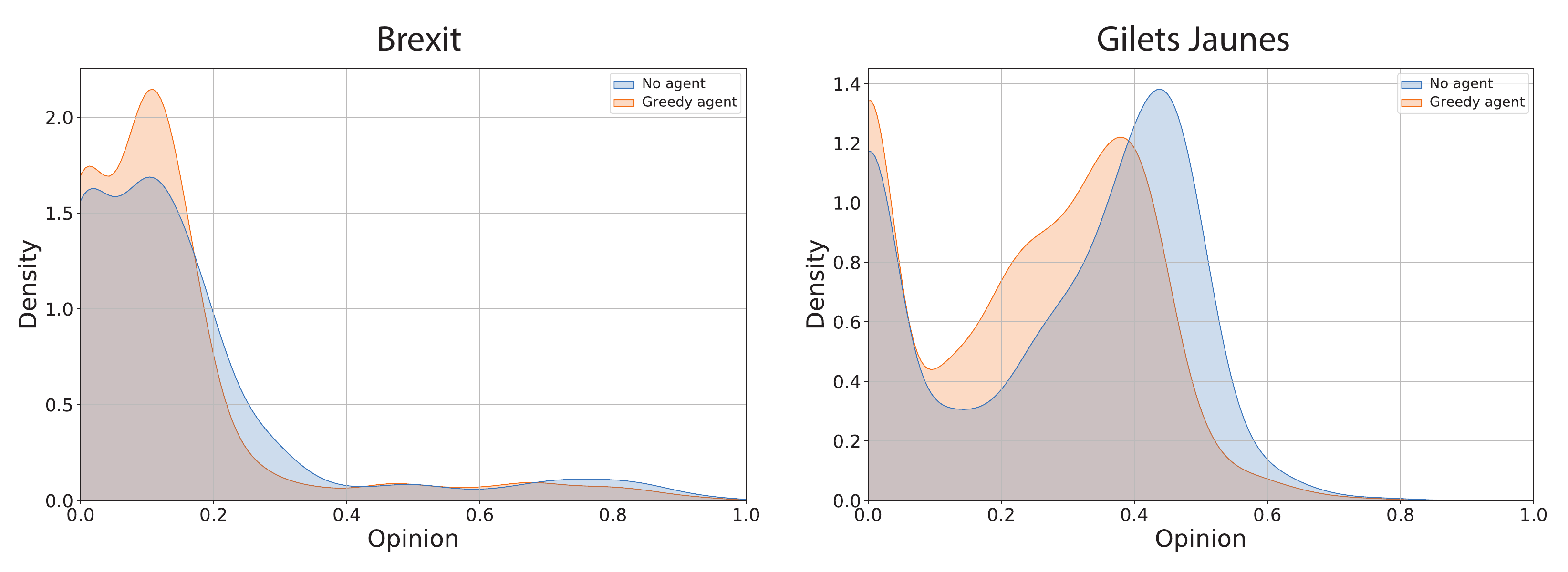}
	\caption{Kernel density estimates for the opinions of non-stubborn nodes in the Brexit and Gilets Jaunes sub-networks with and without the greedy agents connected to their targets. The agents choose targets to minimize the variance.  }
	\label{fig:var_min_opinion_dist} 
\end{figure*}

 To minimize the variance, agent $a_0$ is selected in each iteration.  The algorithm is trying to pull the opinions to zero, as that is where much of the opinions are already located.  By pulling the higher opinions towards the lower values of the majority, the variance is decreased.  This becomes clear by looking at kernel density estimates for the opinions in Figure \ref{fig:var_min_opinion_dist}.  For Brexit the agent we see that the agent increases the density of nodes near 0.1.  In Gilets Jaunes we see that the agent has pulled the mode at 0.43 back to 0.39.  In each dataset, this pull towards zero results in a decreased variance.

 One practical takeaway from the performance on real social networks is that the harmonic influence centrality benchmark is a good compromise between optimization and speed.  Greedy essentially runs a full network harmonic influence centrality calculation for each target it finds.  This can become burdensome for large networks where one searches for hundreds or thousands of targets. 

\section{Opinion Dynamics Model}\label{sec:model}
 We have seen how to use the opinion equilibrium in equation \ref{eq:equilibrium} to optimize functions of the opinions.  We also saw empirical evidence suggesting the equilibrium is a good model for opinions in real social networks in Section \ref{sec:data_stubborn}.  We now provide theoretical support for the use of this equilibrium.
We begin by presenting a general model for the dynamics of opinions between interacting agents in a network.  Our model allows for full heterogeneity among the agents.  There is heterogeneity in the agents' activity levels, meaning the agents can post content at different rates.  There is also heterogeneity in how the agents' opinions evolve in response to seeing new posted content. We show that this model reaches the same equilibrium as the DeGroot model.

\subsection{Model Details}
 Consider the setting of Section \ref{sec:equilibrium} where there are stubborn and non-stubborn agents communicating in a directed network. We will utilize the notation presented in that section.  We now introduce the opinion update rule for the non-stubborn agents. In our analysis, we will focus on a scenario where at each time $t \in \mathbb{Z}_{\geq0}$ a random agent $j\in\mathcal{V}$ communicates with some set of its followers $\mathcal{N}^o_j$ by posting a piece of content. If agent $j$ posts at time $t+1$, we assume that the post has a random opinion $Y_j(t+1) \in [0,1]$ where $\mathbb{E}\bracket{Y_j(t+1)|\theta_j(t)} = \theta_j(t)$. If agent $j$ communicates at time $t+1$ with an agent $i$ such that $j \in \mathcal{N}_i$, then agent $i$ updates his opinion to a convex combination of his own current opinion and agent $j$'s communicated opinion:
\begin{align}\label{eq:update_rule}
\theta_i(t+1) = \left(1-\omega_i(t)   \right) \theta_i(t) + \omega_i(t)Y_j(t+1)
\end{align}
where $\omega_i(t) \in [0,1]$ is some deterministic stubbornness factor for agent $i$ that is changing in time. On the other hand, if agent $i$ does not see a new opinion at time $t+1$, then $\theta_i(t+1) = \theta_i(t)$.

We note that in many previous studies, the random opinion $Y_j(t+1)$ is often always assumed to be agent $j$'s exact opinion at time $t$, given by $\theta_j(t)$. In our analysis, we relax this and only assume that agent $j$ communicates an opinion $Y_j(t+1) \in [0,1]$ that is unbiased, meaning $\mathbb{E}\bracket{Y_j(t+1)|\theta_j(t)} = \theta_j(t)$. This property where the agent does not express his exact opinion in the content he posts is known as limited verbalisation \citep{urbig2003attitude,mason2007situating}.

One should note that if $\omega_i(t)$ shrinks to zero as time increases, agent $i$ weighs communicated opinions less and therefore becomes more stubborn. In previous studies, $\omega_i(t)$ is assumed to be constant in time, which is not necessarily an accurate model of human behavior. As suggested in Mason, Conroy, and Smith \cite{mason2007situating} and Roberts and Viechtbauer \cite{roberts2006patterns}, a model with limited verbalisation and time-evolving update rules is a more realistic model of opinion dynamics.

One interesting case is where $\omega_i(t) = (t+1)^{-1}$ and an agent $i$ observes exactly one opinion at every unit of time.  In this case, it is not difficult to see that 
\begin{align*}
\theta_i(t) = \frac{1}{t}\sum_{s=1}^t Y(s).
\end{align*}
where above we dropped the subscript for the origin of the posts for simplicity. This corresponds to an update rule where an individual's opinion is simply the average of all previous posts he has seen.   

Next we describe the communication pattern of the agents.  We depart from the model in \cite{ghaderi2013opinion}, where at each
discrete time-step all agents in the network communicate.  Rather, we allow the agents to communicate randomly, which is a more accurate model of how individuals in real social networks behave.
Let $p_{ji}$ denote the probability that agent $j$ communicates with agent $i$ at time $t$.  

For our model, we have the following stochastic update rule for non-stubborn agent $i\in \mathcal{V}_1$:
\begin{align*}
\theta_{i}(t+1) = \begin{cases}
\left(1-\omega_i(t)   \right)\theta_i(t) + \omega_i(t)Y_j(t+1) & \quad \text{ w.p. } p_{ji} \text{  if  $j \in \mathcal{N}_i$} \\
\theta_i(t) & \quad \text{ w.p. } 1 - \sum_{j \in \mathcal{N}_i}p_{ji}. \\
\end{cases}
\end{align*}
Taking expectations, we have for non-stubborn agents
\begin{align*}
\mathbb{E} \left[ \theta_i(t+1) \right] = \mathbb{E}\left[\theta_i(t) \right] \left(1 - \omega_i(t)\sum_{j \in \mathcal{N}_i } p_{ji} \right) + \omega_i(t)\sum_{j \in \mathcal{N}_i}\mathbb{E}\left[ \theta_j(t) \right]p_{ji}.
\end{align*}
Then, we can write that 
\begin{align*}
\mathbb E \left[\mathbf{\theta}(t+1)\right] = \mathbf{A} (t) \mathbb E\left[\theta(t)\right]
\end{align*}
where   $\mathbf A(t) = \mathbf I + \mathbf\Omega(t) \mathbf A$,  $\mathbf A$ is  given by equation \eqref{eq:A_matrix}, and $\mathbf\Omega(t)$ is a $| \mathcal V | \times \mathcal V |$ diagonal matrix with $\mathbf \Omega_{ii}(t) = \omega_i(t)$ for non-stubborn agent $i \in \mathcal{V}_1$, and zero otherwise. Throughout, we make the assumption that for all $i\in \mathcal{V}_1$, $\sum_{j\in \mathcal{N}_i} p_{ji} \leq 1$.  
  Finally, to simplify our notation, let $\mathbf \theta_{\mathcal{V}_0}$ denote the vector of the initial opinions of the stubborn agents and $\mathbf \theta_{\mathcal{V}_1}(t)$ denote the vector of the opinions of the non-stubborn agents at time $t$. Similarly, we let $\mathbf \Omega_{\mathcal{V}_1}(t)$ denote the submatrix of $\mathbf{\Omega}(t)$ corresponding to the non-stubborn agents.


\subsection{Theoretical Results}\label{sec:stubborn_agents}
We now present our theoretical results characterizing the opinion equilibrium of the model.  The overarching question is under what conditions on the stubbornness factor $\mathbf \Omega (t)$ do the opinions converge to an equilibrium.  Also, 
are the equilibrium opinions themselves random or do they converge to deterministic values.  
To answer these questions, we have obtained results for the limiting values
of the expectation and variance of the opinions. 
All proofs can be found in Section \ref{sec:proofs}.

We begin with the expectation result.
\begin{theorem}\label{thm:1}
	Suppose the underlying graph $\mathcal{G} \left(\mathcal{V}, \mathcal{E} \right)$ is connected and for each non-stubborn agent $v\in\mathcal V_1$ there exists a directed path from some stubborn agent to $v$. Then, if $\sum\limits_{s=0}^t \min_{i\in \mathcal{V}_1} \omega_i(s)$ diverges we have that
	\begin{align}\label{eq:mean}
	\lim_{t\to \infty} \mathbb E \left[\mathbf \mathbf{\theta}_{\mathcal{V}_1}(t)\right] = -\mathbf G^{-1} \mathbf{F} \mathbf \theta_{\mathcal{V}_0}.
	\end{align}
\end{theorem}

The result states that for convergence in expectation to occur, $\omega_i(t)$ cannot decay too fast.  If this occurs,
then new opinions are ignored and updates will become too small.  This can result
in the final opinion depending upon the initial condition.  However, if $\omega_i(t)$ decays slow enough, where slow means
$\sum_{s=0}^t \min_{i\in\mathcal V_1}\omega_i(s)$ diverges, then the agents will keep listening to new communications and updating their opinions.  In this case, the  expectation of their final opinions are independent of their initial value.    We note that this is the equilibrium expression from equation \eqref{eq:equilibrium}.  Theorem \ref{thm:1} shows that given a network, communication probabilities, and stubborn opinions,  the expectation of the non-stubborn opinions will reach the same equilibrium for a large class of opinion dynamics models.  This provides theoretical support for using this expression when optimizing opinions.


We next consider the variance of the equilibrium opinions. Let $\Sigma\left[ \theta(t) \right] = \mathbb{E} \left[  \left[ \theta(t) - \mathbb{E}[\theta(t)] \right] {\left[ \theta(t) - \mathbb{E}[\theta(t) \right]}]^T   \right]$ denote the covariance matrix of $\theta(t)$. We have the following result.
\begin{theorem}\label{thm:2}
	Suppose the assumptions from Theorem \ref{thm:1} hold. Additionally, suppose that $\sum\limits_{s=0}^t \max_{i \in \mathcal{V}_1}\omega_i^2(s)$ converges. Then, 
	\begin{align*}
	\lim_{t\to \infty} \Sigma\left[ \theta(t) \right] = \mathbf 0.
	\end{align*}
\end{theorem}

Taken together, Theorems \ref{thm:1} and \ref{thm:2} characterize the class of stubbornness factors $\mathbf \Omega(t)$ where convergence occurs in $L^2$. If $\mathbf \Omega(t)$ does not decrease too rapidly ($\sum_{s=0}^t \min_{i\in\mathcal V_1} \omega_i(s)$ diverges) then the expectation of the final opinions of the non-stubborn agents do not depend on their initial conditions. If we also have that $\mathbf \Omega(t)$ decreases sufficiently rapidly ($\sum_{s=0}^t \max_{i\in\mathcal V_1} \omega_i^2(s)$ converges), then the opinions' covariance will go to zero. To parameterize this region, assume $\mathbf \Omega(t)$ has the form  $c_i t^{-\delta_i}$ for some constants $c_i$ and $\delta_i$ for all $i$. Then Theorems \ref{thm:1} and \ref{thm:2} are satisfied for $1/2< \delta_i \leq 1$ for all $i$.


\section{Conclusion}\label{sec:conclusion}
 We have shown here a process for placing stubborn agents in a social network to optimize a variety of functions of the opinions.  The core of this process is an opinion equilibrium condition.  We showed how to evaluate this equilibrium on real social networks using neural networks.
We found that the equilibrium is a good model for opinions in real social networks.  In addition, we provided theoretical support for this opinion equilibrium by proposing a very general class of  models for opinion dynamics in a social network where individuals become more stubborn with time.  We proved that for this very general class of models, this same equilibrium condition is reached.

 Targets for the stubborn agents were identified using a greedy algorithm with the opinions given by the equilibrium.  We were able to establish performance guarantees for this greedy algorithm for maximizing the sum/mean of the opinions due to its monotonicity and submodularity.  We showed that other objectives such as the number of opinions over a threshold and opinion variance did not possess these properties.  Tests on real social networks  showed that the greedy algorithm outperforms several benchmarks, allowing one to obtain greater influence with a limited number of targets for a variety of objective functions. 

 The process developed in this work is a useful operational capability for countering influence campaigns and shaping opinions in large social networks.  As the roles these networks play in our society increases, these types of capabilities will continue to grow in importance.

\bibliography{OR_stubborn}
\bibliographystyle{plainnat}
\ECSwitch

\ECHead{Supplementary Material and Proofs of Statements}

In this E-Companion we provide additional data analysis and technical proofs for the theorems in the paper, ``Optimizing Opinions with Stubborn Agents Under Time-Varying Dynamics.''

\section{Proofs}\label{sec:proofs}
\subsection{Proof of Theorem \ref{thm:centrality}}
For non-stubborn agents, we cannot simply switch their opinions as we did with stubborn agents.  This is because
their opinions depend upon their neighbors opinions. Instead we take a different approach.  To obtain the influence
centrality of a non-stubborn agent $i$,  we connect $d$ stubborn agents to it with communication probability $p$.  We then calculate the change in average opinion
when these stubborn agents' opinion switches from zero to one.  The harmonic influence centrality is given by the limit of this
opinion change as $d$ goes to infinity.  Adding an infinite number of stubborn agents of a single opinion to a non-stubborn agent effectively  makes a non-stubborn agent stubborn.  

We let $\theta_{\mathcal{V}_1}^0$ and $\theta_{\mathcal{V}_1}^1$ correspond to the opinion vector for the non-stubborn agents with the $d$ stubborn agents opinions equal to zero and  one, respectively.  These equilibria are given by
\begin{align*}
\paranth{-\mathbf G + pd\mathbf e_i\mathbf e_i^T} \theta_{\mathcal{V}_1}^0 & =\mathbf F\theta_{\mathcal{V}_0}\\
\paranth{-\mathbf G + pd\mathbf e_i\mathbf e_i^T}  \theta_{\mathcal{V}_1}^1 & =\mathbf F\theta_{\mathcal{V}_0}+pd\mathbf e_i.
\end{align*}
The difference in the equilibrium opinions is given by
\begin{align*}
\theta_{\mathcal{V}_1}^1-\theta_{\mathcal{V}_1}^0 & = \paranth{-\mathbf G + pd\mathbf e_i\mathbf e_i^T}^{-1}\paranth{\mathbf F\theta_{\mathcal{V}_0}+pd\mathbf e_i-\mathbf F\theta_{\mathcal{V}_0}}\\
& = -pd\paranth{\mathbf G - pd\mathbf e_i\mathbf e_i^T}^{-1}\mathbf e_i.
\end{align*}
Because $\mathbf e_i$ is only non-zero in element $i$, we only need to calculate the $i$th column of the inverse
of $\mathbf G - pd\mathbf e_i\mathbf e_i^T$.  This can be done using the Sherman-Morris formula \cite{sherman1950adjustment}, giving
\begin{align*}
\paranth{\mathbf G - pd\mathbf e_i\mathbf e_i^T}^{-1}_{ji}& = G^{-1}_{ji}  + \frac{pdG^{-1}_{ji}G^{-1}_{ii}}{1-pdG^{-1}_{ii}}\\
& = \frac{G^{-1}_{ji}}{1-pdG^{-1}_{ii}}.
\end{align*}
We can now calculate the harmonic influence centrality of a non-stubborn agent.  We must make sure to subtract one, which is the change in the opinion of the given non-stubborn agent.  This is in contrast to stubborn agents, whose opinion shifts were not included
in the calculation of the resulting opinion shift.  With this in mind, and using the above
expression, the harmonic influence centrality of non-stubborn agent $i$ is given by
\begin{align}
c(i) & = \lim_{d\rightarrow\infty}\frac{1}{|\mathcal V_1|-1}\paranth{ \sum_{j\in\mathcal V_1}\paranth{\theta_{\mathcal{V}_1}^1-\theta_{\mathcal{V}_1}^0}_{ji}-1}\nonumber\\  
& = \lim_{d\rightarrow\infty} \frac{1}{|\mathcal V_1|-1} \paranth{\sum_{j\in\mathcal V_1} -pd\paranth{\mathbf G - pd\mathbf e_i\mathbf e_i^T}^{-1}_{ji}-1}\nonumber\\
& =  \lim_{d\rightarrow\infty} \frac{1}{|\mathcal V_1|-1} \paranth{\sum_{j\in\mathcal V_1}\frac{-pdG^{-1}_{ji}}{1-pdG^{-1}_{ii}}-1}\nonumber\\
& = \frac{1}{|\mathcal V_1|-1}\paranth{\frac{\sum_{j\in\mathcal V_1}G^{-1}_{ji}}{G^{-1}_{ii}}-1}\nonumber.
\end{align}

For stubborn agents, we simply change their opinion from zero to one and calculate the change in the mean opinion.
We consider switching the opinion of stubborn agent $i$.  Let $\theta^0$ and $\theta^1$ correspond to the opinion vector with agent $i$'s opinion equal to zero and  one, respectively.

The difference in the equilibrium opinions is given by
\begin{align*}
\theta_{\mathcal{V}_1}^1-\theta_{\mathcal{V}_1}^0 & = -\mathbf G^{-1}\mathbf F\paranth{\theta_{\mathcal{V}_0}^1-\theta_{\mathcal{V}_0}^0}\\
& = -\mathbf G^{-1}\mathbf F\mathbf e_i,
\end{align*}
where $\mathbf e_i$ is a vector of all zeros except for the $i$th component which is equal to one.
We let $c(i)$ be the harmonic influence centrality of stubborn agent $i$ which is equal to the change in the average opinion.  This is then given by
\begin{align}
c(i) & = \frac{1}{|\mathcal V_1|} \sum_{j\in\mathcal V_1}\paranth{\theta_{\mathcal{V}_1}^1-\theta_{\mathcal{V}_1}^0}_{ji}\nonumber\\  
& =\frac{-1}{|\mathcal V_1|} \sum_{j\in\mathcal V_1}\paranth{\mathbf G^{-1}\mathbf F}_{ji}\nonumber.
\end{align}

\subsection{Proof of Theorem \ref{thm:submodular}}
Without loss of generality, we assume the equilibrium condition of the base network is given by
\begin{align*}
-\mathbf G \theta^0 & = \mathbf F\theta_{\mathcal V_0}= \mathbf b.
\end{align*}
We then consider connecting a stubborn edge to arbitrary nodes $k$ and $l$, where the stubborn agent has a posting rate of $p$. Then we have the following different equilibrium conditions:
\begin{align*}
-\mathbf G \theta^0 & = \mathbf b\\
(-\mathbf G +p\mathbf e_k\mathbf e_k^T)\theta^1 & = \mathbf b+p\mathbf e_k\\
(-\mathbf G + p\mathbf e_l\mathbf e_l^T)\theta^2 & = \mathbf b+p\mathbf e_l\\
(-\mathbf G + p\mathbf e_l\mathbf e_l^T + p\mathbf e_k\mathbf e_k^T)\theta^3 & = \mathbf b+p\mathbf e_l + p\mathbf e_k.
\end{align*}
For clarity of notation, we will denote $U(\theta) = \sum_{i\in\mathcal V_1}\theta_i$.  To establish submodularity of the set function $f$, we need to show that
\begin{align*}
U(\theta^1)-U(\theta^0)\geq U(\theta^3)-U(\theta^2).
\end{align*}
We will make use of the following two lemmas, where the proofs of these lemmas are included in the appendix.
\begin{lemma}\label{lemma:delta}
	Let $\theta^0$ be the opinion equilibrium given by $-\mathbf G\theta^0 = \mathbf b$
	and let $\theta^1$ be the opinion equilibrium given by adding a single edge between a stubborn agent and non-stubborn agent $k$ with communication probability $p$.  Then
	\begin{align*}
	U(\theta^1)-U(\theta^0) & = -\frac{p(1 - \theta^0_k)}{1 - pG^{-1}_{kk}}\sum_{i \in\mathcal V_1}G^{-1}_{ki}.
	\end{align*}
\end{lemma}
\begin{lemma}\label{lemma:inverse_positive}
	All elements of the matrix $-\mathbf G^{-1}$ are non-negative.
\end{lemma}
We note that for any $i\in\mathcal V_1$, $\theta^0_i \leq \theta^2_i$ because adding a single stubborn edge with opinion equal to one cannot decrease any non-stubborn opinion.  This follows immediately from Lemmas \ref{lemma:delta} and \ref{lemma:inverse_positive}. Let $\mathbf H = \mathbf G - p\mathbf e_l\mathbf e_l^T$. Using Lemma \ref{lemma:delta} we have
\begin{align}
U(\theta^3)-U(\theta^2) &= -\frac{p(1 - \theta^2_k)}{1 - pH^{-1}_{kk}}\sum_{j\in\mathcal V_1}H^{-1}_{kj}\nonumber\\
&\leq -p(1 - \theta^0_k)\sum_{j\in\mathcal V_1}\frac{H^{-1}_{kj}}{{1 - pH^{-1}_{kk}}}\label{eq:ab_cond}
\end{align}
We now utilize the Sherman-Morrison formula \citep{sherman1950adjustment} which states that
\begin{align*}
(\mathbf G  - p\mathbf e_k\mathbf e_k^T)^{-1}_{ij} & = G^{-1}_{ij} + \frac{pG^{-1}_{ik}G^{-1}_{kj}}{1 - pG^{-1}_{kk}}.
\end{align*}  

Applying the Sherman-Morrison formula to equation \eqref{eq:ab_cond} we obtain
\begin{align*}
U(\theta^3)-U(\theta^2) &\leq -p(1 - \theta^0_k)\sum_{j\in\mathcal V_1}  \frac{G^{-1}_{kj} + \frac{pG^{-1}_{kl}G^{-1}_{lj}}{1 - pG^{-1}_{ll}}}{1 - pG^{-1}_{kk} - p^2\frac{G^{-1}_{kl} G^{-1}_{lk}}{1 - pG^{-1}_{ll}}}\\
&= -p(1 - \theta^0_k)\sum_{j\in\mathcal V_1} \frac{G^{-1}_{kj}(1 -pG^{-1}_{ll}) + pG^{-1}_{kl}G^{-1}_{lj}}
{(1 - pG^{-1}_{kk})(1 - pG^{-1}_{ll})-p^2G^{-1}_{kl}G^{-1}_{lk}} \\
&\leq -p(1 - \theta^0_k)\sum_{j\in\mathcal V_1} \frac{G^{-1}_{kj} }
{(1 - pG^{-1}_{kk})} \\
&=  U(\theta^1)-U(\theta^0)
\end{align*}
where the second inequality follows from the fact that $\mathbf G^{-1}$ is elementwise nonpositive and the final equality follows from Lemma \ref{lemma:delta}, and thus we have now arrived at the desired result.

\subsection{Proof of Theorem \ref{thm:submodular_threshold}}
 This result is proved by constructing a counter-example where the threshold function violates submodularity.  The example network is an undirected path with 50 non-stubborn nodes and two stubborn nodes, illustrated in Figure \ref{fig:path}.  All nodes have equal communication probability.  The stubborn nodes are labeled $s_0,s_1$ and are located at the ends of the path.  The opinions of the stubborn nodes are zero for $s_0$ and one for $s_1$.  The non-stubborn nodes are labeled with integers from 0 to 49, going from left to right in the figure.  Formally, let us define this network as $G=(\mathcal V,\mathcal E)$. The node set is $\mathcal V= \mathcal V_0 \bigcup \mathcal V_1$, $\mathcal V_0 = \curly{i}_{i=0}^{49}$, $\mathcal V_0 = \curly{s_0,s_1}$.  The edge set is $\mathcal E = \curly{(i,i+1)}_{i=0}^{48}\bigcup \curly{(s_0,0),(49,s_1)}$.   The opinion threshold is $\tau = 0.95$.  We use a stubborn agent with opinion one and define $A=\curly{0}, B = \curly{0,1}$, and  $v=6$.  Clearly $A\subset B$.  We solve for the equilibrium opinions with the agent choosing various target sets and show the resulting objective values in Table \ref{table:threshold}. One can check that  $g(A\bigcup v) - g(A) <g(B\bigcup v) - g(B)$, which violates the submodularity condition.



\begin{table}[!hbt] \centering
	\caption{Values for the threshold objective for different target sets  for the network in Figure \ref{fig:path}.  The threshold is $\tau=0.95$ and the agent opinion is one.}
	\label{table:threshold}
	\centering
	\begin{tabular}{|l|c|}
		\hline
		Target set & Number of non-stubborn  \\
		& nodes with opinion $\geq 0.95$\\\hline
		$\curly{0}$ & 5\\\hline
		$\curly{0,6}$ & 33\\\hline
		$\curly{0,1}$ & 12\\\hline
		$\curly{0,1,6}$ & 44\\\hline
	\end{tabular}
\end{table}  
\subsection{Proof of Theorem \ref{thm:variance}}

\emph{Proof.}   As with Theorem \ref{thm:submodular_threshold}, this result can be proven by constructing a suitable counter-example from the network in Figure \ref{fig:path}. We use a stubborn agent $a_0$ with opinion zero and communication probability equal to that of the other nodes in the network. Define the sets $A_i$ for $0\leq i\leq 49$ as 

\begin{align}
    A_i = \curly{(a_0,j)}_{j=0}^i.\label{eq:target_set}
\end{align}
 First we show that the variance is not submodular.  Consider sets $A_0$ and $A_{48}$ and let $v=49$.  We solve for the equilibrium opinions for different target sets and show the non-stubborn opinion variance in Table \ref{table:variance}.  One can check that $A_0\subset A_{48}$ and $h(A_0\bigcup (a_0,v))-h(A_0) < h(A_{48}\bigcup (a_0,v))-h(A_{48})$, which violates the submodularity condition.

 To show that the variance is non-monotone we consider sets $A_{48}$ and $A_{49}$ (note that $A_{49} = A_{48}\bigcup (a_0,49)$).  Using the values in Table \ref{table:variance}, we see that $h(A_{48}) > h(A_{49})$.  Since $A_{48}\subset A_{49}$, this violates the monotonicity condition.



\begin{table}[!hbt] \centering
	\caption{Values for the variance objective for different target sets  for the network in Figure \ref{fig:path}. The target set definitions are given by equation \eqref{eq:target_set}. }
	\label{table:variance}
	\centering
	\begin{tabular}{|l|c|}
		\hline
		Target set & Number of non-stubborn  \\
		& nodes with opinion $\geq 0.95$\\\hline
		$A_0$ & 0.0959\\\hline
		$A_0\bigcup (a_0,49)$ & 0.0429\\\hline
		$A_{48}$ & 0.0430\\\hline
		$A_{48}\bigcup (a_0,49)$ & 0.0386\\\hline
	\end{tabular}
\end{table}  

\subsection{Proof of Theorem \ref{thm:1}}

Recall from the structure of the problem we have that 
\begin{align*}
\mathbb  E \left[\mathbf{\theta}(t+1)\right] &=  \mathbf A (t) \mathbf A (t-1) \cdots \mathbf A (0) \theta(0).
\end{align*}
Due to the structure of the problem, we know that we can write the matrix product in the following block form 
\begin{align*}
\begin{bmatrix}
\mathbf{I}_{|\mathcal{V}_0|\times|\mathcal{V}_0|} & \mathbf{0} \\
\mathbf{F'}(t) &\mathbf{G}'(t)
\end{bmatrix} = \mathbf A (t) \mathbf A (t-1) \cdots \mathbf A (0)
\end{align*}
for some $|\mathcal{V}_1| \times |\mathcal{V}_0|$ matrix $ \mathbf F' (t)$ and some $|\mathcal{V}_1| \times |\mathcal{V}_1|$ matrix $\mathbf{G}'(t)$. Note that the following relations hold for the matrices $\mathbf F'(t)$ and $\mathbf G'(t)$:
\begin{align*}
\mathbf F'(t) &= \mathbf \Omega_{\mathcal{V}_1}(t) \mathbf F + \left[ \mathbf I + \Omega_{\mathcal{V}_1}(t) \mathbf G \right] \mathbf F'(t-1) \\
\mathbf{G}'(t) &= \left[ \mathbf I + \Omega_{\mathcal{V}_1}(t) \mathbf G \right] \mathbf G ' (t-1)
\end{align*}
where $\mathbf{F}'(-1) =\mathbf 0 $ and $\mathbf G ' (-1) = \mathbf I$. Now, because there exists a path to every non-stubborn agent from some stubborn agent, it follows that $\mathbf G^{-1}$ exists. For notation, let $\mathbf E(t) = \mathbf{F}'(t) + \mathbf{G}^{-1} \mathbf F$ and note that
\begin{align*}
\mathbf E (t) &= \mathbf{F}'(t) + \mathbf{G}^{-1} \mathbf F \\
&= \left[ \mathbf I + \mathbf \Omega_{\mathcal{V}_1}(t) \mathbf G \right] \mathbf F ' (t-1) + \left[ \mathbf I + \mathbf \Omega_{\mathcal{V}_1}(t) \mathbf G \right] (\mathbf G^{-1} \mathbf F) \\
&= \left[ \mathbf I + \mathbf \Omega_{\mathcal{V}_1}(t) \mathbf G \right] \mathbf E (t-1) \quad . 
\end{align*}
From $\mathbf E (0) = \mathbf F '(0) + \mathbf{G}^{-1} \mathbf F = \mathbf \Omega_{\mathcal{V}_1}(0) \mathbf F + \mathbf{G}^{-1} \mathbf{F} = \left[ \mathbf I +\mathbf \Omega_{\mathcal{V}_1}(0) \mathbf G \right] \mathbf{G}^{-1}\mathbf F $ we get the relation
\begin{align*}
\mathbf E (t) = \mathbf G'(t) \mathbf{G}^{-1}\mathbf F \quad. 
\end{align*}

We will now make use of the following lemma, which is proved in the  appendix. 
\begin{lemma}\label{lemma:g_convergence}
	If the assumptions in Theorem \ref{thm:1} hold, then $\lim_{t\to \infty}\mathbf{G}'(t) = \mathbf{0} $.
\end{lemma}
From this lemma, we have that $\lim_{t\to \infty} \mathbf E (t) = 0$ and thus $\lim_{t \to \infty}\mathbf  F'(t) = - \mathbf G ^{-1} \mathbf  F$. Therefore, we get that 
\begin{align*}
\lim_{t \to \infty} \mathbb  E \left[\mathbf{\theta}(t)\right] &= \begin{bmatrix}
\mathbf{I}_{|\mathcal{V}_0|\times|\mathcal{V}_0|} & \mathbf{0} \\
-\mathbf G^{-1} \mathbf F & \mathbf 0
\end{bmatrix}  \mathbf \theta (0)
\end{align*}
and thus we have $\lim_{t\to \infty} \mathbb E \left[\mathbf \mathbf{\theta}_{\mathcal{V}_1}(t)\right] = -\mathbf G^{-1} \mathbf{F} \mathbf \theta_{\mathcal{V}_0}$.

\subsection{Proof of Theorem \ref{thm:2}}
We begin by writing $\theta(t+1)$ in the following form:
\begin{align*}
\theta(t+1) = \mathbf A (t) \theta(t) + \mathbf \Omega(t) \epsilon(t)
\end{align*}
where $\epsilon(t)$ is a bounded random vector of dimension $|\mathcal{V}|$ such that $\mathbb{E} \left[ \epsilon(t) \; | \; \theta(t)  \right] = \mathbf{0}$ and $\epsilon_i(t)\in [-1, 1]$. Because of the above relation, we have that 
\begin{align*}
\Cov \left[ \theta(t), \epsilon(t)  \right] &= \mathbb{E} \left[\theta(t)\epsilon(t)^T\right] - \mathbb E \left[\theta(t) \right] \mathbb{E} \left[ \epsilon(t) \right]^T \\
&= \mathbb{E} \left[ \theta(t)\mathbb{E} \left[  \epsilon(t)^T \; | \; \theta(t) \right]   \right] - \mathbb E \left[\theta(t) \right] \mathbf{0}^T \\
&= \mathbb{E} \left[ \theta(t)\mathbf{0}^T \right] \\
&= \mathbf{0} .
\end{align*}
Due to the fact that $\epsilon(t)$ and $\theta(t)$ are uncorrelated, we have the following:
\begin{align*}
\Sigma\left[  \theta(t+1) \right] &= \mathbf{A} (t)\Sigma\left[ \theta(t) \right] \mathbf{A}(t)^T + \mathbf \Omega (t) \Sigma\left[ \epsilon(t) \right] \mathbf \Omega (t). 
\end{align*}
Using the fact that $\Sigma \left[ \theta(0) \right] = \mathbf{0}$ and that the stubborn agents' opinion vector is constant in time, we have that
\begin{align*}
\Sigma\left[  \theta_{\mathcal{V}_1}(t+1) \right]  &= \sum\limits_{j=0}^{t} \overline{\mathbf{B}}(t,j+1)\Sigma \left[ \overline{\epsilon}(j) \right]{\overline{\mathbf{B}}(t, j+1)}^T 
\end{align*}
where $\overline{\epsilon}(t)$ is the subvector of $\epsilon(t)$ corresponding to the non-stubborn agents and
\begin{align*}
\overline{\mathbf{B}} (t, j+1) =\left( \mathbf I + \mathbf \Omega_{\mathcal{V}_1}(t) \mathbf G \right) \cdots \left( \mathbf I + \mathbf \Omega_{\mathcal{V}_1}(j+1) \mathbf G \right) \mathbf \Omega_{\mathcal{V}_1} (j) 
\end{align*}
for $j+1 \leq t$ and $\overline{\mathbf{B}} (t, j+1) = \Omega_{\mathcal{V}_1} (j)$ for $j+1>t$.

From the triangle inequality and the non-negativity of matrix norms we now have that 
\begin{align*}
\norm{\Sigma\left[  \theta_{\mathcal{V}_1}(t+1) \right]}_{\infty} &\leq  \sum\limits_{j=0}^{\infty} \norm{\overline{\mathbf{B}}(t,j+1)\Sigma \left[ \overline{\epsilon}(j) \right]{\overline{\mathbf{B}}(t, j+1)}^T }_{\infty}\\ 
&\leq \sum\limits_{j=0}^{\infty} \norm{\overline{\mathbf{B}}(t,j+1) }_{\infty} \norm{ \Sigma \left[ \overline{\epsilon}(j) \right] }_{\infty} \norm{{\overline{\mathbf{B}}(t, j+1)}^T }_{\infty} 
\end{align*}
where the second inequality follows from the submultiplicativity of the maximum absolute row sum norm. Now, from Popoviciu's inequality we know that $\Var\left\{\epsilon_i(t)\right\} \leq 1$ for all $i$. Furthermore, by Cauchy-Schwarz we know that $ | \Cov\bracket{ \epsilon_i(t), \; \epsilon_j(t)}|  \leq 1$ for all $i$ and $j$. From this, we know that $\norm{ \Sigma \left[ \overline{\epsilon}(j) \right] }_{\infty} \leq |\mathcal{V}_1 |$ for all $j$. 

Furthermore, because $\left( \mathbf I + \mathbf \Omega_{\mathcal{V}_1}(i) \mathbf G \right)$ is substochastic for all $i$ we know that $\left( \mathbf I + \mathbf \Omega_{\mathcal{V}_1}(t) \mathbf G \right) \left( \mathbf I + \mathbf \Omega_{\mathcal{V}_1}(t-1) \mathbf G \right) \cdots \left( \mathbf I + \mathbf \Omega_{\mathcal{V}_1}(j+1) \mathbf G \right)$ is substochastic for all $t$ and $j$. From this we know that $\norm{\overline{\mathbf{B}}(t,j+1) }_{\infty} \leq \max_{i \in \mathcal{V}_1} \omega_i(j)$ and $\norm{\overline{\mathbf{B}}(t,j+1)^T }_{\infty} \leq |\mathcal{V}_1| \max_{i \in \mathcal{V}_1} \omega_i(j)$. Combining all of these bounds, we have that
\begin{align*}
\norm{\overline{\mathbf{B}}(t,j+1) }_{\infty} \norm{ \Sigma \left[ \overline{\epsilon}(j) \right] }_{\infty} \norm{{\overline{\mathbf{B}}(t, j+1)}^T }_{\infty} \leq |\mathcal{V}_1|^2 \max_{i \in \mathcal{V}_i} \omega_i(j)^2 .
\end{align*}
From the assumptions in the Theorem, $\sum_{j}\max_{i \in \mathcal{V}_1} \omega_i(j)^2$ converges and thus from the dominated convergence theorem we have that 
\begin{align*}
\lim_{t\to\infty} \norm{\Sigma\left[  \theta_{\mathcal{V}_1}(t+1) \right]}_{\infty} &\leq \sum\limits_{j=0}^{\infty} \lim_{t\to\infty} \norm{\overline{\mathbf{B}}(t,j+1)\Sigma \left[ \overline{\epsilon}(j) \right]{\overline{\mathbf{B}}(t, j+1)}^T }_{\infty}
\end{align*}
It follows easily from a variant of Lemma \ref{lemma:g_convergence} that $\lim_{t \to \infty}\overline{\mathbf{B}}(t,j+1) = \mathbf 0 $ for all $j$, and thus we have arrived at the desired result $\lim_{t\to\infty} \Sigma\left[  \theta_{\mathcal{V}_1}(t+1) \right]=0$.

\subsection{Proof of Lemma \ref{lemma:delta}}
We assume the matrix $\mathbf G$ is $n\times n$.
We assume we have an equilibrium solutions $-\mathbf G \theta^0 = \mathbf b$
and $-(\mathbf G - p\mathbf e_k\mathbf e_k^T)\theta^1= \mathbf b +p\mathbf e_k$.  Using the Sherman-Morrison formula, we can write the difference in objective functions as
\begin{align*}
U(\theta^1)-U(\theta^0) & = \sum_{i=1}^n{\left({-\left(\mathbf G - p\mathbf e_k\mathbf e_k^T\right)}^{-1}(\mathbf b+p\mathbf e_k)\right)}_i + \sum_{i=1}^n (\mathbf G^{-1}\mathbf b)_i\\
& = -p\sum_{i=1}^n (\mathbf G^{-1}\mathbf e_k )_i - p\sum_{i=1}^n {\left(\frac{\mathbf G^{-1}\mathbf e_k\mathbf e_k^T\mathbf G^{-1}}{1 - pG^{-1}_{kk}}{\left(\mathbf b+p\mathbf e_k\right)}\right)}_{i}\\
& = -p\sum_{i=1}^n\paranth{ G^{-1}_{ki} + \frac{G^{-1}_{ki}(-\theta^0_k+pG^{-1}_{kk})} {1 - pG^{-1}_{kk}}}\\
& = -\frac{p(1 - \theta^0_{k})}{1 - pG^{-1}_{kk}}\sum_{i=1}^n G^{-1}_{ki}.
\end{align*}

\subsection{Proof of Lemma \ref{lemma:inverse_positive}}
First off, note that the matrix $-\mathbf{G}$ is a Z-matrix because all of its off-diagonal elements are non-positive. Furthermore,the eigenvalues of  $-\mathbf{G}$ have positive real part by Gershgorin's Circle theorem \citep{gershgorin1931uber} and the fact that $-\sum_{j=1}^n G_{ij}\geq 0$. Thus, by definition, $-\mathbf{G}$ is an M-matrix. The result follows from the fact that the inverse of a non-singular M-matrix is nonnegative. See \cite{plemmons1977m} for a review of Z and M-matrices, and for the details on the proof of the fact that the inverse of a non-singular M-matrix is non-negative.


\subsection{Proof of Lemma \ref{lemma:g_convergence}}
We begin by defining a matrix sequence that is somewhat similar to $\mathbf{G}'(t)$, given by 
\begin{align*}
\mathbf G ''(t) = \left(\mathbf I + \mathbf G \min_{i\in \mathcal{V}_1} \omega_i(t) \right) \cdots \left(\mathbf I + \mathbf G \min_{i\in \mathcal{V}_1} \omega_i(0) \right).
\end{align*}

For clarity, we will break the proof of this lemma into the following two parts:
\begin{enumerate}
	\item We begin by showing that $\norm{ \mathbf G'(t) } _{\infty} \leq \norm{\mathbf G''(t) }_{\infty}$ for all $t$, where $\norm{\cdot}_{\infty}$ is the maximum absolute row sum norm of a matrix.
	\item We then show that $\lim\limits_{t \to \infty} \mathbf G ''(t) = \mathbf 0 $.
\end{enumerate}
Note that by combining the two parts we will have arrived at the desired result by the continuity of matrix norms. We now present the proof of the above two parts.
\subsubsection{Proof of 1.}

First, note that $\mathbf G'(t)$ is a nonnegative substochastic matrix for all $t$. For notation, let $\mathbf e$ denote the vector of all ones. We first show that $\mathbf G\mathbf G'(t) \mathbf e\leq 0 $ for all $t$. Note that this holds for $t=0$ because 
\begin{align*}
\mathbf G \mathbf G'(0) \mathbf e &= \mathbf G \mathbf e + \mathbf G \mathbf \Omega(0) \mathbf G \mathbf e \quad .
\end{align*}
Now, by construction $\mathbf G$ is a matrix with negative diagonal entries that are bounded below by $-1$, and all other entries non-negative. Additionally, for notation let $\mathbf G \mathbf e = \mathbf c \leq 0$. Then we have that $\mathbf G \mathbf G'(0) \mathbf e = \mathbf c + \mathbf G \mathbf \Omega (0) \mathbf c \leq \mathbf c - \mathbf c = \mathbf 0$ since $\mathbf \Omega(0)$ is a diagonal matrix with entries between 0 and 1. We now show the result for general $t$ by induction. For clarity, let $\mathbf G \mathbf G'(t) \mathbf e = \mathbf{c}_t$, and by the inductive assumption $\mathbf{c}_t \leq \mathbf{0} $. Then we have that 
\begin{align*}
\mathbf{G}\mathbf{G}'(t+1)\mathbf e &= \mathbf G \left[  \mathbf I + \mathbf \Omega_{\mathcal{V}_1} (t+1)\mathbf G \right] \mathbf G'(t) \mathbf e \\
&= \mathbf c_t + \mathbf G  \mathbf \Omega_{\mathcal{V}_1} (t+1)\mathbf c_t \\
&\leq \mathbf c_t - \mathbf c_t \quad . 
\end{align*}
And thus the result holds for general $t$.

Now, using this we show that
\begin{align*}
\mathbf G'(t)  \mathbf e \leq  \mathbf G ''(t)  \mathbf e
\end{align*}
for all $t$. Note that the above trivally holds for $t=0$ because, by construction $\mathbf G \mathbf e \leq 0$. Now, once again we prove the result by induction. We have that 
\begin{align*}
\mathbf G '(t+1) \mathbf e = \mathbf{G}'(t) \mathbf e + \mathbf \Omega_{\mathcal{V}_1} (t+1)\mathbf G \mathbf G'(t) \mathbf e
\end{align*}
and from above we know that $\mathbf G \mathbf G'(t) \mathbf e \leq 0$. Thus we have that 
\begin{align*}
\mathbf G '(t+1) \mathbf e \leq \left( \mathbf I + \mathbf G \min_{i\in \mathcal{V}_1} \omega_i(t+1) \right) \mathbf G'(t) \mathbf e 
\end{align*}
Then, from the inductive assumption and the fact that $\mathbf I + \mathbf G \min_{i\in \mathcal{V}_1} \omega_i(t+1)$ is a non-negative matrix we have that
\begin{align*}
\mathbf G'(t+1)  \mathbf e \leq \mathbf G''(t+1) \mathbf e 
\end{align*}

Once again, because $\mathbf G'(t)$ is a nonnegative matrix, the above result implies that 
\begin{align*}
\norm{ \mathbf G'(t) } _{\infty} \leq \norm{ \left(\mathbf I + \mathbf G \min_{i\in \mathcal{V}_1} \omega_i(t+1) \right) \cdots \left(\mathbf I + \mathbf G \min_{i\in \mathcal{V}_1} \omega_i(0) \right) }_{\infty}
\end{align*}
where $\norm{\cdot}_{\infty}$ is the maximum absolute row sum norm of a matrix.

\subsubsection{Proof of 2.}

Now, let $\lambda$ be an arbitrary eigenvalue of $\mathbf A$ with corresponding eigenvector $v$. Then, due to the fact that $\mathbf I + \mathbf A$ is a stochastic matrix, by the Perron-Frobenius Theorem we know that $|\lambda + 1|\leq 1$ for all eigenvalues $\lambda$ of $\mathbf A$. Now, consider the Jordan canonical form of the matrix $\mathbf A$. First off, because $\mathbf I + \mathbf A$ is stochastic and reducible with aperiodic recurrent states, then the algebraic and geometric multiplicities of the eigenvalue equal to one (of $\mathbf I + \mathbf A$) are equal to the number of communication classes \cite[p.~198]{bremaud2013markov}. Because of this, we know that the Jordan Canonical form of $\mathbf A$ can be written as 
\begin{align*}
\mathbf A = \mathbf V \begin{bmatrix} \mathbf 0_{|\mathcal V_0| \times |\mathcal V_0|} & & & \\
& \mathbf J_1 & & \\
& & \ddots & \\
& & & \mathbf J_k \end{bmatrix} \mathbf V^{-1}
\end{align*}
where  $\mathbf J_i$ are Jordan block matrices of the form 
\begin{align*}
\mathbf J_i = \begin{bmatrix} \lambda_i & 1 & & &\\
& \lambda_i &\ddots & &\\
& & & \ddots & 1\\
& & & & \lambda_i
\end{bmatrix} 
\end{align*}
and $|1+\lambda_i| < 1$ for all $i \in \left\{1, \ldots, k \right\}$. Furthermore, the matrix $\mathbf V$ has as columns the respective generalized eigenvectors of $\mathbf A$. By expressing the Jordan matrix as $\mathbf J$, we have the standard Jordan canonical form $\mathbf A = \mathbf V \mathbf J \mathbf V^{-1}$. Consider this generalized eigenvector matrix $\mathbf V$, and for notation we denote $\mathbf V$ in the following block matrix form:
\begin{align*}
\mathbf V  = \begin{bmatrix} \mathbf V_{00} & \mathbf V_{01} \\
\mathbf V_{10} & \mathbf V_{11} \end{bmatrix}
\end{align*}
where $\mathbf V_{00}$ is a $|\mathcal{V}_0| \times |\mathcal{V}_0|$ matrix and $\mathbf V_{11}$ is a $|\mathcal{V}_1| \times |\mathcal{V}_1|$ matrix. We first show that $\mathbf V_{01} = \mathbf 0$. First, let $v_{i,m}$ be a generalized eigenvector associated with $\lambda_i$ of rank $m$ where $|1+\lambda_i| < 1$. We thus know that 
\begin{align*}
{\left(A - \lambda_i \mathbf I \right)}^m v_{i,m} = \mathbf 0 \implies (-\lambda_i)^m v_{i,m}(j) = 0
\end{align*}
for all $j \in \mathcal{V}_0$, where $v_{i,m}(j)$ is the $j$-th entry of the vector $v_{i,m}$. From this, we know that $\mathbf V_{01} = \mathbf 0$. Thus, because $\mathbf V^{-1}$ exists we now know that $\mathbf V_{11}^{-1}$ exists and furthermore, we have that 
\begin{align*}
\mathbf{V}^{-1} = \begin{bmatrix}
\mathbf{V}_{00}^{-1} & \mathbf 0 \\
- \mathbf{V}_{11}^{-1} \mathbf V _{10} \mathbf V_{00}^{-1} & \mathbf V_{11}^{-1}
\end{bmatrix}
\end{align*}
from the block matrix inversion formula  \cite[p.~44]{bernstein2005matrix}. Therefore, from $\mathbf A = \mathbf V \mathbf J \mathbf V^{-1}$ we have that $\mathbf{G} = \mathbf V_{11} \mathbf J' \mathbf{V}_{11}^{-1}$ where $\mathbf J'$ is the Jordan matrix made up of the Jordan blocks $\mathbf J_i$ for $i \in \left\{1, \ldots, k \right\}$.

Now, consider the function given by 
\begin{align*}
f_t(\mathbf  G) &= \left(\mathbf I + \mathbf G \min_{i\in \mathcal{V}_1} \omega_i(t) \right) \cdots \left(\mathbf I + \mathbf G \min_{i\in \mathcal{V}_1} \omega_i(0) \right) \\
&= \mathbf V_{11} f_t(\mathbf J') \mathbf V_{11}^{-1}.
\end{align*}

Due to the block structure of the Jordan matrix $\mathbf J$, we have that 
\begin{align*}
f_t(\mathbf J') = \begin{bmatrix} 
f_t\left( \mathbf J_1 \right) & & \\
& \ddots & \\
& &  f_t\left( \mathbf J_k\right) \end{bmatrix} \quad. 
\end{align*}
For every Jordan block matrix $\mathbf J_i$ we will now show that $\lim\limits_{t\to \infty} f_t\left(\mathbf J_i\right) = \mathbf 0 $.

For notation let $D = \left\{ z \in \mathbb C : |z+1|<1  \right\}$, and let $n_i\times n_i$ be the size of the Jordan block $\mathbf J_i$.  Because the function $f_t$ is a product of analytic functions on $D$, then $f_t$ is analytic on the set $D$. Thus we have that
\begin{align*}
f_t\left( \mathbf J_i \right) = \begin{bmatrix} f_t(\lambda_i) & f_t^{(1)}(\lambda_i) & \frac{f_t^{(2)}(\lambda_i)}{2!} & \cdots & \frac{f_t^{(n_i-1)}(\lambda_i)}{(n_i-1)!} \\ 
& f_t(\lambda_i) & f_t^{(1)}(\lambda_i) & \cdots & \frac{f_t^{(n_i-2)}(\lambda_i)}{(n_i-2)!} \\
& & \ddots & \ddots& \vdots \\ 
& & & f_t(\lambda_i)& f_t^{(1)}(\lambda_i)\\
& & & & f_t(\lambda_i)
\end{bmatrix}
\end{align*}
where $f_t^{(j)}$ is the $j$-th derivative of the function $f_t$ (see \cite[p.~2-4]{higham2008functions} for more information on analytic functions applied to the Jordan canonical form of a matrix).  

Now, because $\sum \limits_{j=0}^t \min_{i\in \mathcal{V}_1}\omega_i(j)$ diverges, this implies that the sequence of functions $\left\{ f_t\right\}_{t\geq0}$ converges uniformly to 0 on any $E \subset D$ such that $E$ is a closed subset of $\mathbb C$. For $x \in D$ take $r_{x} > 0$ such that $\left\{ z: |z-x| \leq r_x \right\} \subset D$. Then, from Cauchy's Differentiation formula we have that 
\begin{align*}
f_t^{(j)}(x) = \frac{j!}{2\pi i} \int_{|z-x| = r_x} \frac{f_t(z)} { (z-x)^{j+1} }dz \quad .
\end{align*}
Now, we have that $|f_t(z)| \to 0$ uniformly on $\left\{z:|z-x| = r_x \right\}$. Thus, from the above $f_t^{(j)}(z) \to 0$. 
By applying the above to every eigenvalue $\lambda_i \in D$, we obtain the desired result.

\section{Dataset Construction}\label{sec:data_analysis}

For Brexit, we used the Twitter Stream API from 09/27/2018 to 01/31/2019. We only used the keyword ``Brexit'' (case-insensitive), which brought us 26,947,305 tweets and 2,131,724 users. We then selected a subset of 104,755 users, who were those who posted at least three tweets mentioning Brexit during the first two weeks of data collection and we collected their user profiles. 

For the Gilets Jaunes protest, we used a variety of keywords to search for tweets, which are shown in Table \ref{table:hashtag_choices_YellowVests}. As can be seen, this set of keywords captures a much broader set of tweets than tweets only about Gilets Jaunes. The reason is that Gilets Jaunes by itself is not discussed enough, hence we would not have enough tweets to train the neural network. We streamed from 01/26/2019 to 04/29/2019 and recovered 3.2 million tweets posted by 370,210 users. We then selected a subset of approximately 40,000 users that appeared to be active on the Gilets Jaunes topic. We looked at users who used any of the keywords also in their description, and collected their full timelines (according to the Twitter search API documentation, we can only recover the last 3,200 tweets posted by each user). We then filtered out the tweets that did not contain any of the keywords in Table \ref{table:hashtag_choices_YellowVests}, which left us with a set of 198,000 new tweets. We added this to the streamed database to get our final dataset. 

The keywords used to label the neural network training data for the Brexit and Gilets Jaunes datasets are shown in Tables \ref{table:hashtag_choices_BREXIT} and \ref{table:hashtag_choices_YellowVests}.  The strings that do not correspond to words are unicode sequences for different images known as emojis.  For instance, the emoji in Table \ref{table:hashtag_choices_BREXIT} corresponds to the flag of the European Union.

\begin{table}[h!]
	\begin{center}
		\caption{Keywords used for construction of the neural network training data for Brexit.}
		\label{table:hashtag_choices_BREXIT}
		\begin{tabular}{|l|l|} 
			\hline
			\multicolumn{1}{|c|}{Pro-Brexit}   & \multicolumn{1}{|c|}{Anti-Brexit}  \\
			\hline
			BrexitmeansBrexit, SupportBrexit , & FBPE, StopBrexit, \\
			HardBrexit,Full Brexit, & strongerin, greenerin,\\
			LeaveMeansLeave, & intogether,\\
			Brexiter, Brexiteer, & infor, remain, \\ 
			antieu, Anti EU, & Bremain,  votein, \\
			no2eu,   wtobrexit, & incrowd, \\
			FullBrexitProperExit & yes2europe, \\
			ProBrexit, PlanAPlus, & exitfrombrexit,   Eunity, \\
			ChuckCheques, ChuckCheq,&  Forthemany,  DeeplyUnhelpful,\\
			voteleave, votedleave,&  WATON,  ABTV,\\
			ivotedleave, &  EUsupergirl,\\
			voteout,	votedout, &  FBSI, NHSLove, \\
			pro-brexit,	pro brexit, probrexit, &   U0001f1eaU0001f1fa  \\
			takebackcontrol,	betteroffout, &\\
			StandUp4Brexit, WeAreLeaving &  \\
			\hline
			
		\end{tabular}
	\end{center}
\end{table}

\begin{table}[h!]
	\begin{center}
		\caption{Keywords used for the construction of the neural network training data for Gilets Jaunes.  These keywords were also used to collect tweets to build the Gilets Jaunes dataset.}
		\label{table:hashtag_choices_YellowVests}
		\begin{tabular}{|l|l|l|} 
			\hline
			\multicolumn{1}{|c|}{Pro-Gilets Jaunes}   & \multicolumn{1}{|c|}{Anti-Gilets Jaunes}  & \multicolumn{1}{|c|}{Mixed} \\
			\hline
			YellowVests, & giletsbleu, & GrandD\'ebat, \\
			violencepoliciere, & cr\'etinsjaunes, & GrandDebat,\\
			\'EtatDeDroit, & cretinsjaunes, & EmmanuelMacron, \\
			EtatDeDroit,  & STOP\c caSuffit,& Macron\\
			r\'epression, & stopcasuffit, &\\
			\'etatpolicier, & TouchePasAMonEurope,&\\
			EtatPolicier, & CetteFoisJeVote,&\\
			 Anti EU, & EnsembleavecMacron,&\\
			Acte16, Acte17, Acte18,  & SoutienAuPr\'esidentMacron,&\\
			 MacronDemission, Frexit & SoutienAuPresidentMacron,&\\
			 & U0001f1eaU0001f1fa, U0001F1EBU0001F1F7,&\\
			 & U0001F1EBU0001F1F7, U0001f1eaU0001f1fa &\\
			\hline
			
		\end{tabular}
	\end{center}
\end{table}
\section{Neural Network}\label{sec:neural_network}
To asses the opinion of a tweet, we used a convolutional neural network architecture. Each tweet is first preprocessed in two versions and sent to two channels in the neural network. The model architecture  was inspired by  \cite{kim2014convolutional}.  Their approach was to train a text classification model on two different word embeddings of the same text: one static channel comprised of embeddings using word2vec \citep{goldberg2014word2vec} and another channel which is the output of an embedding layer.

\begin{figure}[h!]
		\centering
	\includegraphics[scale=0.5]{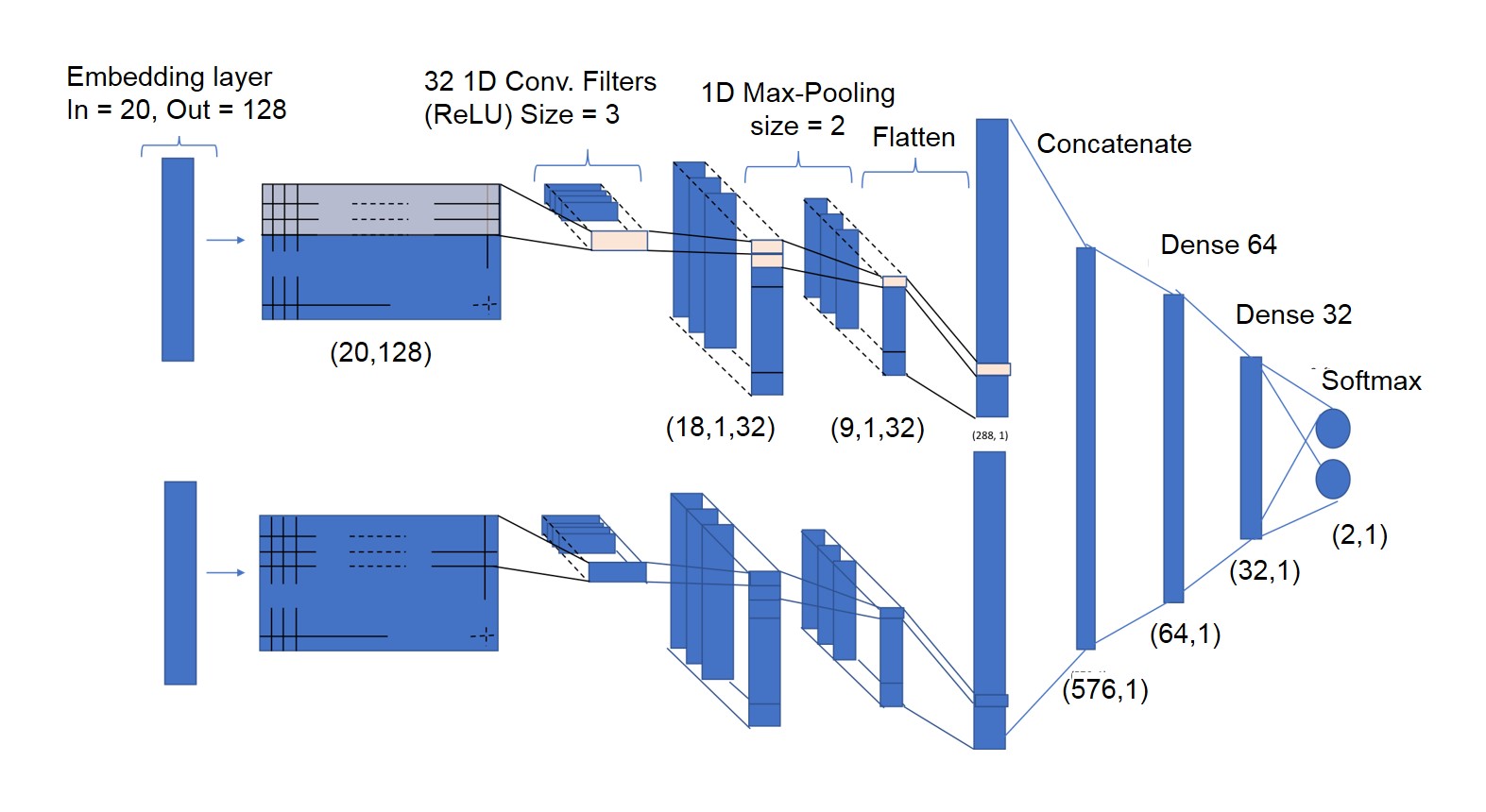}
	\caption{Diagram of the neural network architecture used to learn tweet opinions.  }
	\label{fig:CNN}
\end{figure}

We show the complete neural network architecture in  Figure \ref{fig:CNN}.
Each tweet is pre-processed into two one-hot encodings (see Section \ref{sec:preprocessing}).
Then, each version of the processed tweet goes through its own embedding layer \textit{(dimension dense embedding = 128)} that will then output two separate channels, each of size (20, 128). Each channel will go through its own separate 32 1D-convolution filters \textit{(kernel size = 3, stride = 1, padding = ‘valid’)}. Convolution filters enable one to represent n-grams  and learn shared parameters by convolving on various parts of the tweet. This prevents overfitting and enables one to learn translation invariant features. We then use a ReLU activation which is know to provide nice gradients for optimization and alleviate the problem of vanishing gradients. After the activation, we implement 1D max-pooling layers (pool size = 2). Pooling enables one to reduce computational cost, and enhance translational invariance by focusing on parts of the input where signals are the strongest. After pooling, we  use a flattening layer. The resulting output is two (288,1) layers that we concatenate to form a (576,1) layer. This layer then goes through two fully connected layers with a ReLU activation and 64 and 32 units, respectively. The final layer is a softmax layer that outputs the probability of being equal to one.

\subsection{Data Pre-processing}\label{sec:preprocessing}
Before being used to train the neural network, each tweet goes through a processing phase where we remove punctuation and stopwords and convert it into a format that the network can process. Each processed tweet is then converted  into two versions.
One version keeps hashtags as they are. This results in a one-hot encoding  vector of size $|\mathcal{V}|$, where $\mathcal{V}$ is the vocabulary of words when hashtags are left as they are.
The second encoding splits hashtags into actual words. This  results in a one-hot encoding of size $|\mathcal{V}^*|$, where $\mathcal{V}^*$ is the vocabulary of words when hashtags are broken down into separate words.

For example, \textit{I hope @candidate\_x will be our next president \#voteforcandidate\_x \#hatersgonnahate.} will be converted into two versions: 
\begin{itemize}
	\item[-] \textit{I hope candidate\_x will be our next president voteforcandidate\_x hatersgonnahate} 
	\item[-] \textit{I hope candidate\_x will be our next president vote for candidate\_x haters gonna hate}.
\end{itemize}
We  do this in order to prevent the neural network from being a lazy learner which only learns from the hashtags. 
This can also bring in more information since words are usually built on roots. For example, the commonly used hashtag  \emph{\#standUpForBrexit} will be broken down into stand + up + for + br + exit, hence  conveying the idea of exit as a good thing. If a new tweet is posted and mentions \emph{the necessary exit from the EU} then it will receive a score closer to pro-Brexit.

The hashtag splitting was done using the WordSegment library in Python \citep{wordsegment}. The sequence length of the tweets was set to 20 tokens (i.e. words). Any tweet with more than 20 tokens is truncated, while tweets with less than 20 tokens are padded with zeros.

\end{document}